\documentclass{article}

\PassOptionsToPackage{numbers, compress}{natbib}
\usepackage[preprint]{neurips_2024}

\usepackage[utf8]{inputenc} 
\usepackage[T1]{fontenc}    
\usepackage{hyperref}       
\usepackage{url}            
\usepackage{booktabs}       
\usepackage{amsfonts}       
\usepackage{nicefrac}       
\usepackage{microtype}      
\usepackage{xcolor}         
\usepackage{enumitem} 
\usepackage{graphicx}
\usepackage{listings}
\usepackage{floatflt}
\usepackage{wrapfig}
\usepackage{amsmath}
\usepackage{multirow}
\usepackage{subcaption}
\usepackage[most]{tcolorbox}
\usepackage{inconsolata}
\usepackage{float}

\title{OmniScientist: Toward a Co-evolving Ecosystem of Human and AI Scientists}

\author{
\textbf{Chenyang Shao}$^{1,2}$\hspace{3mm}
\textbf{Dehao Huang}$^2$\hspace{3mm}
\textbf{Yu Li}$^1$\hspace{3mm} 
\textbf{Keyu Zhao}$^1$\hspace{3mm} 
\textbf{Weiquan Lin}$^2$\hspace{3mm} \\
\textbf{Yining Zhang}$^2$\hspace{3mm} 
\textbf{Qingbin Zeng}$^{1}$\hspace{3mm}
\textbf{Zhiyu Chen}$^2$\hspace{3mm} 
\textbf{Tianxing Li}$^1$\hspace{3mm}
\textbf{Yifei Huang}$^2$\hspace{3mm} \\
\textbf{Taozhong Wu}$^{2}$\hspace{3mm} 
\textbf{Xinyang Liu}$^1$\hspace{3mm}  
\textbf{Ruotong Zhao}$^1$\hspace{3mm}  
\textbf{Mengsheng Zhao}$^2$\hspace{3mm} 
\textbf{Jiaoyang Li}$^2$\hspace{3mm} \\
\textbf{Xuhua Zhang}$^{2}$\hspace{3mm}  
\textbf{Yue Wang}$^{2}$\hspace{3mm} 
\textbf{Yuanyi Zhen}$^{2}$\hspace{3mm}
\textbf{Fengli Xu}$^{1,2,*}$\hspace{3mm} 
\textbf{Yong Li}$^{1,2,*}$\hspace{3mm}
\textbf{Tie-Yan Liu}$^{2}$\hspace{3mm}
\\
$^1$Department of Electronic Engineering, BNRist, Tsinghua University\\
$^2$Zhongguancun Academy \\
$^*$\{fenglixu, liyong07\}@tsinghua.edu.cn
\vspace{1mm}\\
\parbox{0.1\textwidth}
{\includegraphics[width=\linewidth]{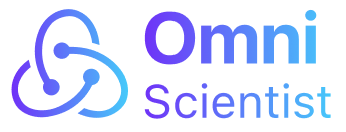}}\hspace{0.5mm}\textcolor{cyan}{\nolinkurl{omniscientist.ai}}\vspace{-3mm}
}

\begin{document}

\maketitle
\begin{abstract} 

With the rapid development of Large Language Models (LLMs), AI agents have demonstrated increasing proficiency in scientific tasks, ranging from hypothesis generation and experimental design to manuscript writing. 
Such agent systems are commonly referred to as ``AI Scientists.'' 
However, existing AI Scientists predominantly formulate scientific discovery as a standalone search or optimization problem, overlooking the fact that scientific research is inherently a social and collaborative endeavor. 
Real-world science relies on a complex scientific infrastructure composed of collaborative mechanisms, contribution attribution, peer review, and structured scientific knowledge networks. 
Due to the lack of modeling for these critical dimensions, current systems struggle to establish a genuine research ecosystem or interact deeply with the human scientific community.

To bridge this gap, we introduce \textbf{OmniScientist}, a framework that explicitly encodes the underlying mechanisms of human research into the AI scientific workflow. 
\textbf{OmniScientist} not only achieves end-to-end automation across data foundation, literature review, research ideation, experiment automation, scientific writing, and peer review, but also provides comprehensive infrastructural support by simulating the human scientific system, comprising: (1) a structured knowledge system built upon citation networks and conceptual correlations; (2) a collaborative research protocol (\textbf{OSP}), which enables seamless multi-agent collaboration and human researcher participation; and (3) an open evaluation platform (\textbf{ScienceArena}) based on blind pairwise user voting and Elo rankings. 
This infrastructure empowers agents to not only comprehend and leverage human knowledge systems but also to collaborate and co-evolve, fostering a sustainable and scalable innovation ecosystem.
Through \textbf{OmniScientist}, we aim to transition AI agents from mere task executors to genuine scientists capable of understanding scientific norms, participating in collaboration, and driving the evolution of the scientific ecosystem. 
\end{abstract}

\newpage

\begingroup
\renewcommand{\baselinestretch}{1}\normalsize
\tableofcontents
\endgroup

\newpage

\section{Introduction}

The practice of science has always evolved with its tools, from the telescope and the microscope to the computer and the algorithm.
Today, large language models (LLMs) represent the next major transformation.
Across disciplines, LLM-powered agents are beginning to assist in tasks once reserved for human researchers: reviewing vast literatures, proposing hypotheses, writing reports, and even designing experiments.
As these capabilities deepen, a fundamental question arises: Can AI evolve from a mere tool into a genuine participant in the scientific ecosystem?

Current efforts to build such ``AI Scientists''~\cite{Lu2024TheAS, Tang2025AIResearcherAS} have made significant strides. Systems like AlphaEvolve~\cite{Novikov2025AlphaEvolveAC} perform iterative optimization through explicit mathematical modeling and code-
based exploration of search spaces, while OpenAI Deep Research~\cite{openai_deep_research_2025} conducts broad information
retrieval and synthesis guided by a specified research topic.
Systems such as Virtual Lab~\cite{Swanson2024TheVL} and
Future House~\cite{futurehouse_platform} represent a further step toward automation, integrating more comprehensive AI-
driven research workflows and coordinating multiple tools to accomplish complex scientific tasks.
However, despite their sophistication, these approaches predominantly formulate scientific discovery as a standalone search or optimization problem, overlooking a fundamental reality: scientific research is inherently a social and collaborative endeavor supported by a complex institutional infrastructure.
Due to the absence of these critical dimensions,
current systems operate as isolated tools, struggling to establish a genuine research ecosystem or to interact deeply with the human scientific community.

Integrating the human research infrastructure is essential for advancing AI scientific intelligence.
Centuries of scientific progress have yielded not just static facts, but a sophisticated cognitive and structural framework. For instance, citation networks transform isolated findings into a traceable lineage of ideas, revealing the evolutionary path of scientific thought; peer-review mechanisms act as rigorous quality controls ensuring reliability; and collaborative protocols regulate the exchange of contribution and credit. 
These structures provide the necessary ``environment'' for science to evolve. 
Without explicitly modeling and encoding these underlying mechanisms, AI scientists remain efficient executors but fail to inherit the dynamic, self-correcting nature of human scientific research.

In this paper, we take the first step by introducing \textbf{OmniScientist}\footnote{\url{omniscientist.ai}}, a comprehensive framework that explicitly encodes human research infrastructure into the life cycle of AI-driven research. 
OmniScientist goes beyond simple task automation; it simulates a complete scientific environment.
At its core lies a robust \textbf{data foundation}, built upon millions of full-text publications and metadata. This forms a dynamic scientific network that captures citation relationships and knowledge context, serving as the system's cognitive bedrock. 
Building on this foundation, the \textbf{literature review} module employs a multi-agent architecture to conduct iterative, semantically guided exploration, ensuring that agents possess a comprehensive awareness of the research landscape. 
Guided by this context, the \textbf{research ideation} process leverages principles from the science of science~\cite{Fortunato2018ScienceOS} to explore and refine concepts within the citation network, generating novel hypotheses that are both contextually grounded and methodologically rigorous. 
For \textbf{experiment automation}, the system employs an iterative multi-agent loop that generates, evaluates, and refines experimental strategies, enabling self-optimization through rigorous feedback mechanisms. 
Following experimentation, \textbf{scientific writing} is supported by an integrated framework that synthesizes related work, generates figures, and refines text according to standardized academic norms, producing coherent, publication-ready manuscripts. 
Finally, the system incorporates a \textbf{paper review} mechanism that functions as a quality control gate, evaluating submissions through in-depth comparison with prior work to provide objective and actionable feedback. Together, these components do not function in isolation but as an interconnected ecosystem covering the entire lineage of scientific research.

\begin{figure}[h]
    \centering
    \includegraphics[width=1\linewidth]{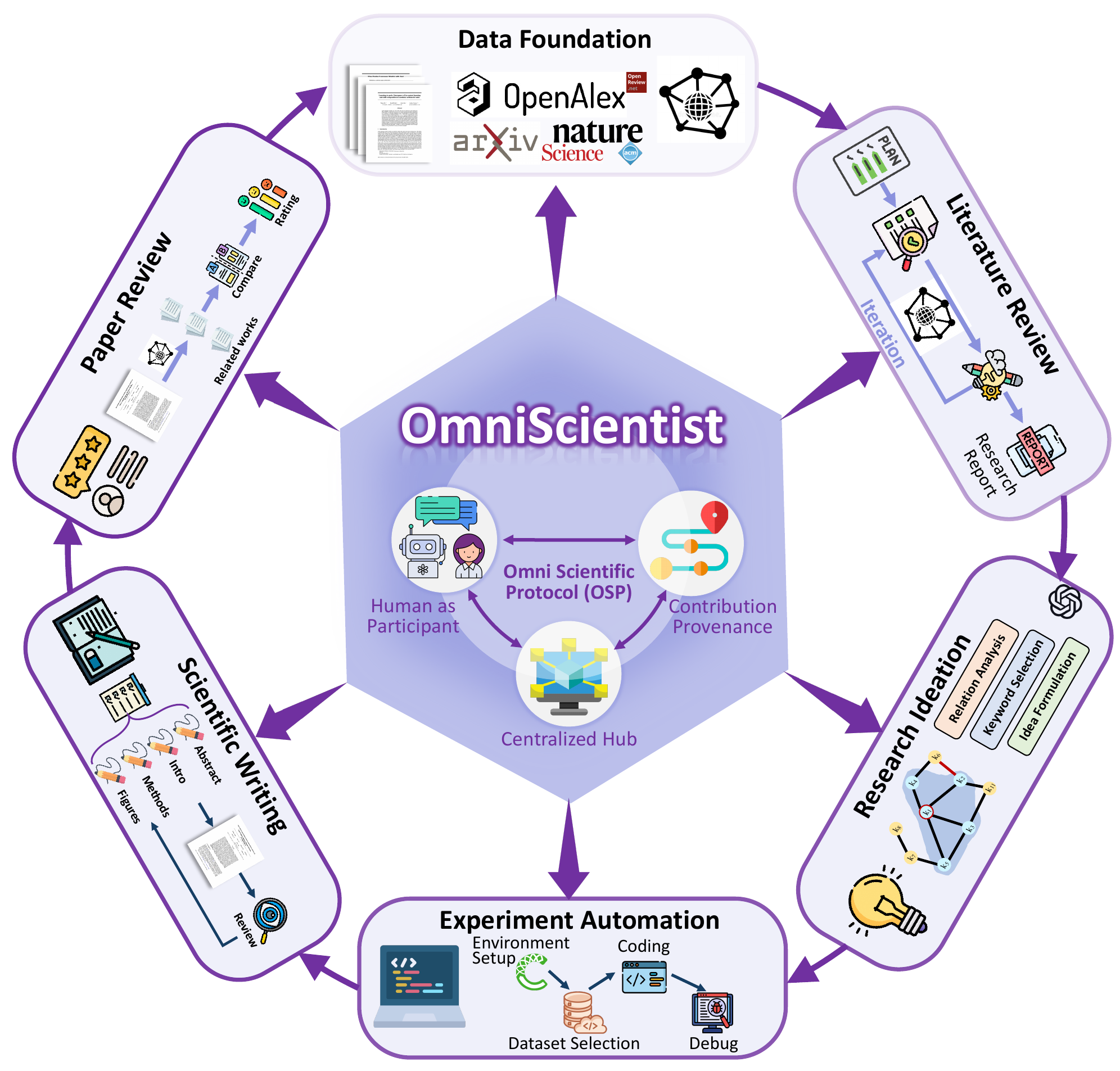}
    \caption{Overview of Our OmniScientist System}
    \label{fig:placeholder}
\end{figure}

Furthermore, to transform these functional modules into a cohesive and governed ecosystem, we introduce two critical infrastructural innovations. 
First, we propose the \textbf{Omni Scientific Protocol (OSP)}, a standardized collaboration backbone designed to orchestrate the complex interactions between multiple AI agents and human researchers. 
Rather than treating humans as passive observers, OSP allows researchers to seamlessly participate in the execution and collaboration processes, providing timely feedback, strategic suggestions, or course corrections whenever the system requires high-level human intuition. 
To maintain scientific integrity, OSP further incorporates a granular contribution tracking system. This mechanism records the provenance of every idea, dataset, and experimental result, attributing credit to specific agents or human participants, thereby establishing a transparent model of authorship and accountability akin to the contributor roles in modern science. 
Second, addressing the persistent challenge of evaluating open-ended scientific discovery, we develop \textbf{ScienceArena}\footnote{\url{ScienceArena.ai}}, an open benchmarking platform designed to simulate the community-driven nature of scientific validation. 
Unlike static metrics, ScienceArena employs a blind, pairwise voting mechanism where human experts evaluate anonymized research outputs based on scientific rigor and novelty. 
By aggregating these preferences into dynamic Elo ratings, the platform establishes a living leaderboard that reflects evolving community standards, effectively allowing human judgments to actively shape the evolutionary direction of AI scientific agents.

Collectively, this work marks a paradigm shift from designing isolated research tools to constructing a comprehensive scientific ecosystem. 
By explicitly encoding the infrastructure of human research into the AI workflow, OmniScientist empowers LLM agents to evolve from mere task executors into autonomous participants within this community. 
Looking ahead, we envision a future where AI Scientists autonomously refine their capabilities through continuous evolution within the ecosystem, while collaborating with human researchers to collectively expand the boundaries of knowledge.

\section{Research Scope}

An \textbf{AI Scientist} refers to a system that employs artificial intelligence, particularly LLMs, to emulate scientific research activities. 
Its objective is to perform various stages of scientific investigation with minimal human intervention, including generating novel hypotheses, designing experimental protocols, conducting experiments or simulations, analyzing results, and even drafting research manuscripts. 
In essence, AI Scientists aim to endow AI with the creativity and reasoning capabilities of human scientists, enabling autonomous or semi-autonomous scientific discovery. 
AI Scientists represent a significant shift in the role of AI, moving from a computational tool toward an originator of scientific knowledge.

A typical AI Scientist system consists of two main components: \textit{intelligent planning} and \textit{automated execution}. 
The \textit{planning component} is often implemented with LLMs, serving as the system's cognitive core. It excels at extracting knowledge from literature and data, performing reasoning, and planning experimental procedures. The \textit{execution component} operates as the system's hands and instruments, using code, simulation tools, or robotic platforms to carry out experiments and collect data according to the planned procedures. 
This design can form a closed loop in which the AI Scientist repeatedly executes the core cycle of the scientific method: generating hypotheses, conducting experiments, measuring outcomes, and refining the hypotheses. 
This loop allows the system to continuously learn and improve from feedback, analogous to human scientists iteratively testing and refining their theories.

In recent years, multiple systems and platforms worldwide have explored the AI Scientist vision, with technical approaches converging into three main directions. 
These directions collectively reveal the evolution of AI research from automated execution to method-driven reasoning and reflect differing perspectives on whether AI can serve as an independent agent of scientific discovery.

The first direction emphasizes \textbf{fully automated workflows}. Representative systems include the AI Scientist~\cite{Lu2024TheAS} developed by Sakana AI in Japan and Westlake University's DeepScientist~\cite{weng2025deepscientist}.
Sakana AI's AI Scientist is among the earliest systems to claim fully automated scientific discovery. The system uses a LLM as its central control unit, orchestrating tasks across different stages through predefined templates. In 2025, its AI Scientist version 2~\cite{yamada2025ai} achieved a notable milestone: three papers generated autonomously by the AI were submitted to an ICLR Workshop, and one of these was even accepted. 
DeepScientist advances the discovery-driven research paradigm by formalizing scientific discovery as a Bayesian optimization problem, enabling the AI to refine its actions through multi-level experimental loops. 
The system has achieved SOTA performance in several AI-related tasks, demonstrating the feasibility of machine-led scientific research, although its applications remain largely confined to computational and simulated environments rather than experimental natural sciences.
Overall, these systems demonstrate that a closed-loop AI research process is achievable, yet they operate largely under a constrained and predefined task context.

The second direction involves \textbf{human-AI collaborative paradigms}, exemplified by DeepMind's AI Co-Scientist~\cite{gottweis2025towards}. Unlike fully automated systems, this approach emphasizes complementary collaboration between AI and human scientists. The system employs multiple dedicated agents, each fulfilling a distinct cognitive role in the research process, such as hypothesis generation, critical evaluation, ranking, and evolutionary optimization. Through mechanisms such as Elo scoring and iterative feedback, the agents collectively form a socially structured scientific reasoning process, simulating interactions in real research teams. 
Experiments show that this group intelligence can propose testable hypotheses in complex domains such as biomedical research, some of which have been published in leading journals. The work demonstrates the potential of human-AI collaboration to enhance discovery.

The third direction encompasses \textbf{knowledge-augmented research platforms}, represented by FutureHouse's AI research ecosystem~\cite{futurehouse_platform} and DPTechnology's Bohrium platform~\cite{bohrium_platform}. FutureHouse emphasizes open interfaces and a modular agent architecture, including agents such as Crow for literature assistance, Falcon for review synthesis, Owl for novelty assessment, and Phoenix for experimental planning. 
These agents collaborate in a pipeline to help researchers identify key knowledge, detect innovation opportunities, and generate executable experimental plans from extensive literature. The platform integrates open databases and domain-specific corpora, ensuring high transparency and traceability in research knowledge retrieval, analysis, and validation.
Bohrium focuses on cross-stage integration through the concept of a Science Navigator, unifying literature comprehension, computational simulation, and experimental execution into a single research operating system. 
The platform can answer natural language scientific queries, perform simulations, and execute experiments, creating a closed loop between information processing and experimental workflows. 
These systems extend AI capabilities from textual understanding to scientific computation and experimental interfacing, illustrating the potential for AI Scientists to integrate with real research facilities. 
Current applications are mainly in materials science and chemistry, but the architectural concepts provide a reference for future AI research clouds.

In summary, existing AI Scientist systems mark a crucial transition from static tools to more autonomous research agents. They are increasingly capable of undertaking laborious research workflows across expansive digital and physical environments.
However, most current approaches still predominantly formulate scientific discovery as a standalone search or optimization problem. They often operate without the support of the foundational infrastructure that sustains human scientific research. Our OmniScientist fills this critical gap by explicitly encoding these underlying mechanisms into the AI workflow. By embedding these structural dimensions, OmniScientist advances the paradigm from merely optimizing research efficiency to constructing a scientific ecosystem, enabling AI agents to navigate rigorous scientific norms and evolve from standalone executors into autonomous participants in the scientific discovery.

\section{Key Designs}

\subsection{Data Foundation}
\label{data_foundation}

We construct a dynamic and comprehensive research knowledge base that not only supports the full spectrum of scientific activities, ranging from literature review to ideation, but also mirrors the collaborative fabric of human scientific ecosystems. 

First, we incorporate the OpenAlex open-access academic graph, one of the most comprehensive scholarly knowledge networks. The dataset contains approximately 269 million paper metadata records, along with their citation relationships. For each paper, the metadata fields include, but are not limited to: title, abstract, authors, affiliations, publication year, venue, DOI, references, citation counts, keywords, subject area, and open-access status. This structured information provides a foundational map of the scholarly landscape, encoding not just academic content but also the web of attribution, influence, and topic lineage critical to scientific collaboration.

Second, we integrate the arXiv open-access paper repository, providing approximately 2.6 million PDF full-text documents, covering over 90\% of AI-related publications. This resource supports deep semantic reading and content-based reasoning beyond metadata, empowering the system to perform actions similar to those of human researchers conducting comprehensive literature surveys.

In addition, to capture actionable scientific outputs and experimental artifacts, we collect 102,679 full-text papers from the top ten AI conferences over the past decade, along with 116,970 referenced baseline models and 68,316 related datasets. Each entry encapsulates not only the textual description of methods and results but also implementation resources, datasets, hyperparameters, and evaluation metrics. This structure enables the system to trace scientific workflows, reproduce experiments, and explore methodological innovations.

\begin{figure}
    \centering
    \includegraphics[width=\textwidth]{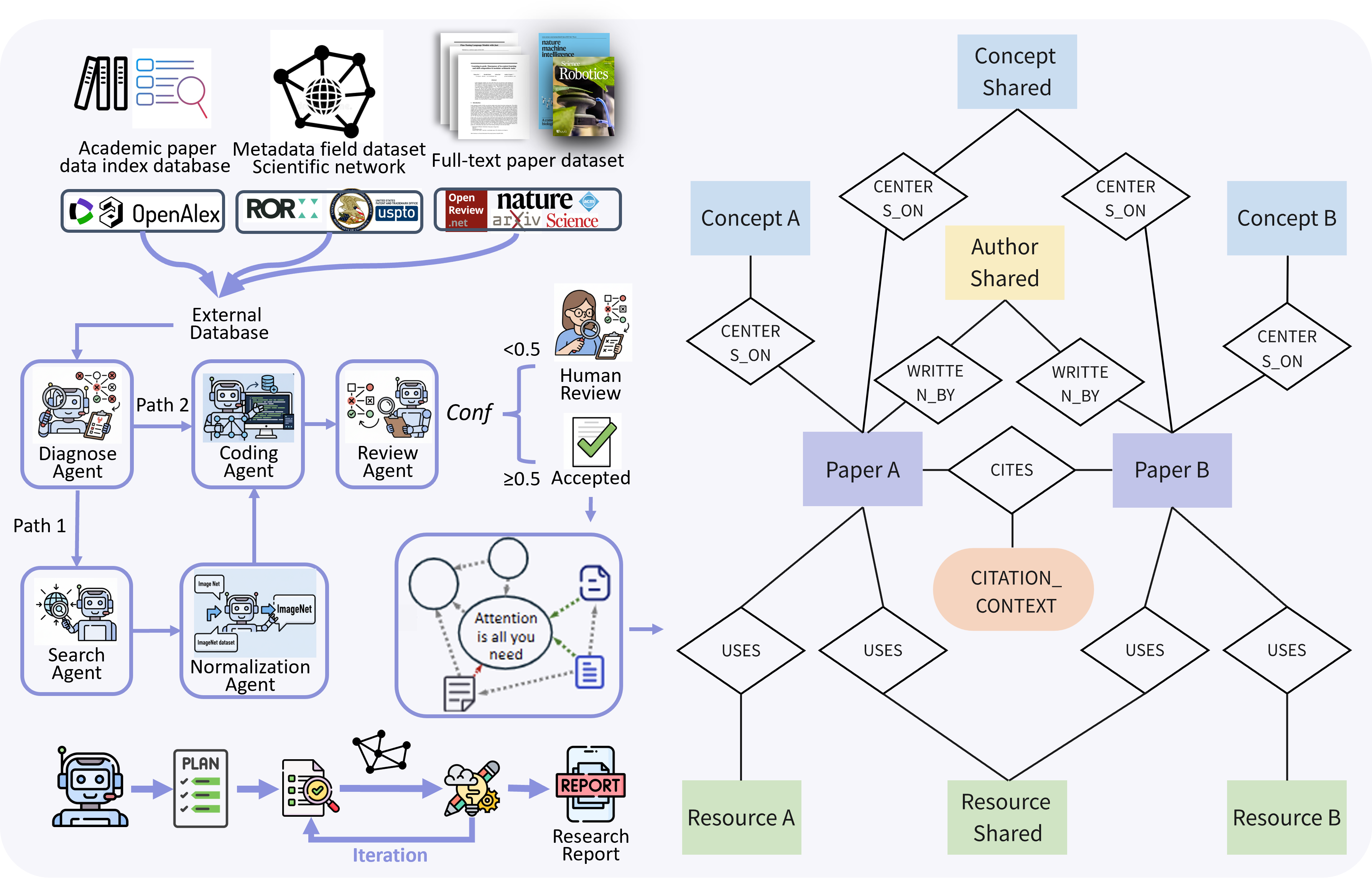}
    \caption{The Multi-Agent Refinement Pipeline (left) and the Refined Data Structure (right).}
    \label{fig:pipeline}
\end{figure}

The knowledge base is organized as a directed, labeled graph comprising four core node types: Paper, Author, Concept, and Resource (datasets, models, tools). The semantic structure is captured through edges such as CITES (Paper to Paper), WRITTEN\_BY (Paper to Author), USES (Paper to Resource), and CENTERS\_ON (Paper to Concept). As shown in Figure~\ref{fig:pipeline} (right), this graph schema represents not only data but also the underlying structure of scientific progress. To further model the interpretive layer of scientific discourse, we attach citation\_contexts to CITES edges, preserving the textual rationale behind citations. This allows the system to move beyond structural links to reason over authors' intents and comparative judgments.

However, scientific knowledge is not static. Like human ecosystems that rely on editorial boards, peer review, and collaborative curation, our data foundation must remain dynamic, self-improving, and structurally coherent. We therefore deploy a multi-agent refinement pipeline (Figure~\ref{fig:pipeline}, left) that continuously diagnoses, enriches, and validates the graph. 

\begin{itemize} 
\item \textbf{Diagnose Agent}: This agent initiates the cycle by auditing the current state of the knowledge graph for quality issues. It prioritizes these issues and formulates refinement tasks, effectively creating a to-do list of data curation actions. 

\item \textbf{Search Agent}: Once issues are identified, the Search Agent queries external scholarly databases and APIs and infers the right answer. It also parse full-text papers to extract hidden semantic metadata: e.g., detecting if a paper mentions using a particular dataset or codebase, or if it cite another work positively or negatively. 

\item \textbf{Normalization Agent}: The agent standardizes the result of the Search Agent to ensure that identical entities are not duplicated under different names (e.g., “ImageNet dataset” vs. “Image Net”). 

\item \textbf{Coding Agent}: The Coding Agent takes the normalized information and integrates it into the knowledge graph. It acts as the database editor, merging duplicate entities and inserting new nodes or relations in alignment with the graph schema. 

\item \textbf{ReviewAgent}: The final agent in the pipeline acts as a quality control and validation layer. The Review Agent evaluates the modifications made to the knowledge graph in the current iteration, checking for accuracy and coherence. If certain new edges are found to be erroneous or low-confidence, the Review Agent can remove them and store them for human review.
\end{itemize}

\begin{table}
    \centering
    \begin{tabular}{|c|c|c|}
        \hline
        \textbf{Metric} & \textbf{Original OpenAlex} & \textbf{Refined Database} \\
        \hline
        Metadata Completeness Score & 0.965 & 1.000 \\
        \hline
        Metadata Correctness Score & 0.951 & 0.997 \\
        \hline
        Retrieval Accuracy (QA Benchmark) & 0.700 & 0.880 \\
        \hline
    \end{tabular}
    \caption{Performance Metrics of the Multi-Agent Refinement Pipeline}
    \label{tab:performance_metrics}
\end{table}

\textbf{Performance Metrics.} We conducted a small-batch evaluation (n = 1000) to assess improvements in structure and retrieval. Metadata completeness increased from 0.965 to 1.000, and correctness from 0.951 to 0.997. On a benchmark of 100 QA pairs probing inter-paper relationships, retrieval accuracy improved from 0.70 to 0.88. These gains reflect the system's enhanced ability to surface relational evidence critical for knowledge reuse and synthesis.

\begin{figure}
    \centering
    \includegraphics[width=1.0\textwidth]{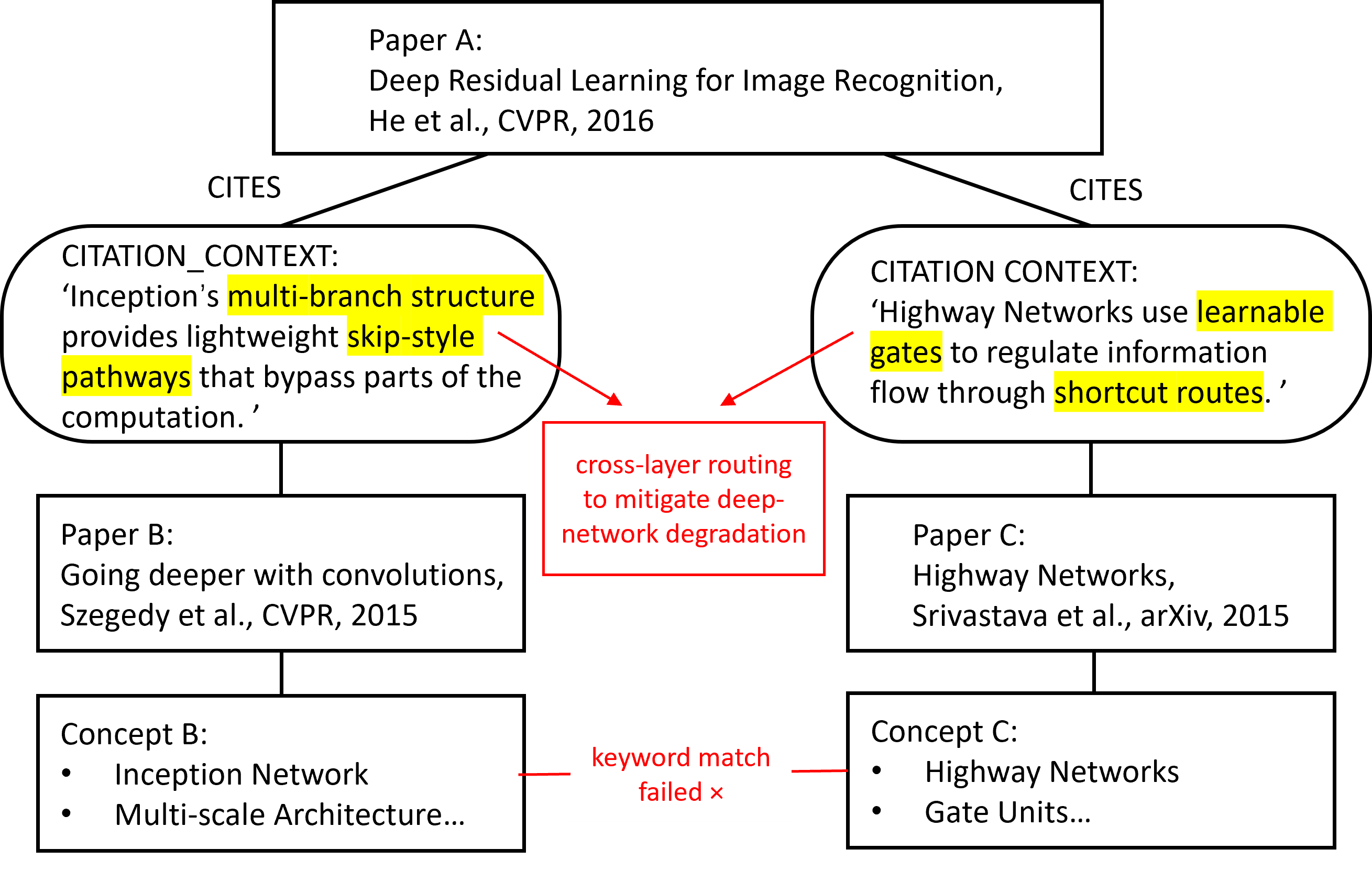}
    \caption{Case Study: Semantic Relation Capture}
    \label{fig:case_study}
\end{figure}
\textbf{Case Study: Semantic Relation Capture.} 

Figure~\ref{fig:case_study} illustrates how the refined scientific network uncovers subtle, author-intended conceptual links that traditional metadata or keyword searches cannot reveal. In this example, Paper A cites Paper B and Paper C through two distinct CITES relations. Viewed independently, B and C appear unrelated: they address different topics, share no metadata, and do not cite each other. Under conventional keyword retrieval, they remain in separate semantic regions.

The refined KG exposes a deeper alignment. The citation\_context from A to B emphasizes that Inception's multi-branch structure provides lightweight, skip-style pathways that bypass parts of the computation. The citation\_context from A to C highlights that Highway Networks use learnable gates to regulate information flow through shortcut routes. Considered together, these contexts show that both works investigate mechanisms for routing information across layers to mitigate degradation in deep models; one achieves this through fixed architectural branches, while the other employs parameterized gating. They also reveal a methodological contrast: Paper B treats shortcut pathways as architectural design, whereas Paper C frames them as trainable control mechanisms. By representing citation contexts as structured relational properties, the refined knowledge graph makes this hidden conceptual bridge explicit, significantly enhancing the granularity of the literature review agent.

Together, the structured knowledge base and refinement pipeline form a dynamic substrate for agent collaboration, scientific inference, and cross-agent memory. By mirroring the protocols and epistemic structures of human scientific research, our data foundation sets the stage for an AI research ecosystem capable of cumulative innovation and sustained interaction with the human scientific community.

\subsection{Literature Review}

Current general-purpose literature review products such as OpenAI DeepResearch exhibit several limitations when applied to scientific research contexts. First, they suffer from low-quality information sources. The open web contains heterogeneous and noisy information with limited factual verification; such content may include erroneous, misleading, or overly speculative claims that are incompatible with the rigor, accuracy, and norms required for scientific inquiry. Second, their retrieval depth is shallow. 
These systems predominantly rely on keyword-based or embedding-based semantic search, which is insufficient for conducting complex scientific literature retrieval. Third, they demonstrate limited scientific rigor and weak reasoning discipline. As these products were not originally designed for research workflows, their summaries and inferences often fail to meet the standards of precision, structure, and methodological soundness expected in scholarly practice.

To address these issues, OmniScientist introduces a semantically grounded and rigor-oriented automated literature review pipeline specifically designed for scientific research. 
To mitigate the problem of low-quality information sources, we construct a local scientific paper database together with curated scientific networks (as detailed in Section~\ref{data_foundation}), ensuring that the review relies only on verified, peer-reviewed research outputs.
To improve the precision and completeness of retrieval, we build an Elasticsearch service on top of the local database, enabling multi-field querying across titles, abstracts, author metadata, and other structured fields.
More importantly, we leverage the constructed \textbf{scientific network} to further enhance both the breadth and depth of retrieval. Specifically, the initial set of candidate papers is obtained from the embedding-based Elasticsearch search. Papers linked through citation and reference relationships in the scientific network are then added to the candidate pool. A relevance verification mechanism is applied to filter this pool, and the retained papers are further expanded layer by layer along their citation and reference links. 
\textbf{This process goes beyond keyword matching to emulate how human researchers trace the ``genealogy of ideas,''} effectively implementing a breadth-first search (BFS) that navigates the structural infrastructure of the field until a predefined retrieval depth or coverage is reached. This network-augmented retrieval process is implemented as a flexible tool, allowing a balance to be struck between retrieval depth and efficiency.


Starting from a user-defined research query, the literature review system executes retrieval, filtering, parsing, and synthesis through coordinated multi-agent orchestration. The detailed workflow is as follows:

\begin{figure}
    \centering
    \includegraphics[width=0.9\textwidth]{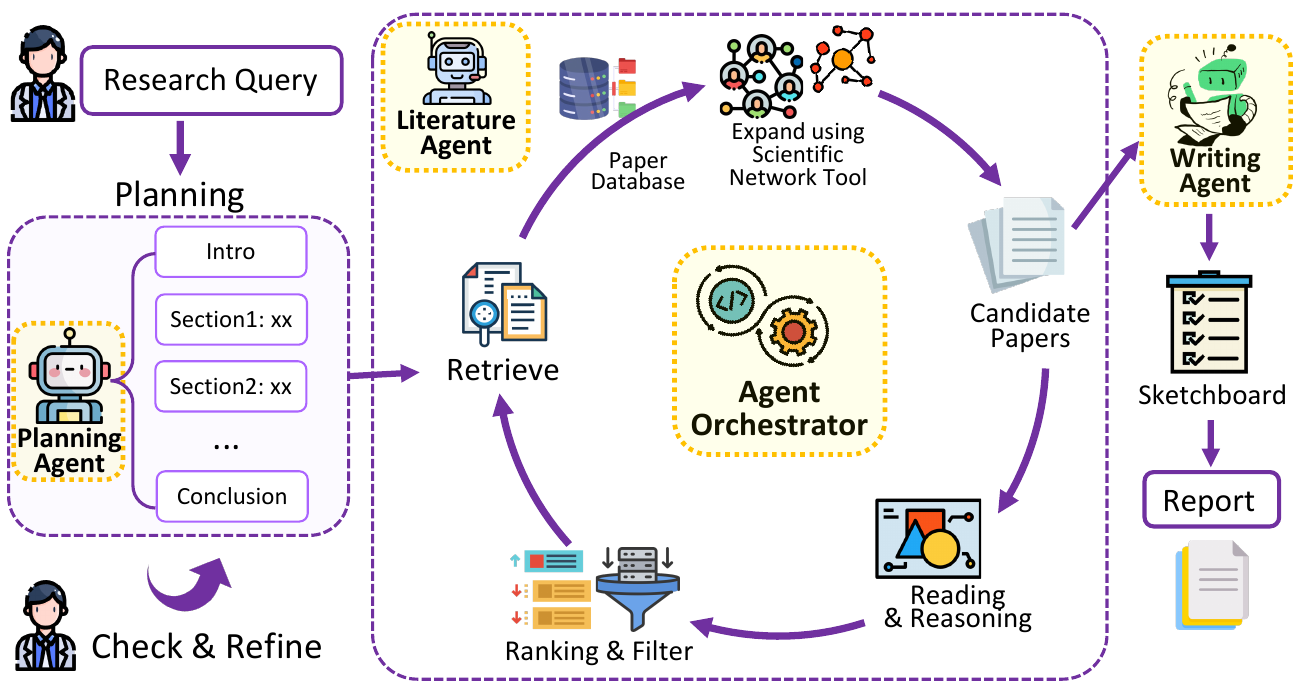}
    \caption{Deep Research Framework Diagram}
    \label{fig:placeholder}
\end{figure}

\begin{itemize} 
    \item \textbf{Research Plan Generation.}  
    Upon receiving a research topic or problem statement, the Planning Agent synthesizes an initial research plan, identifying research objectives, decomposed sub-questions, expected methodological directions, and key conceptual dimensions. These elements serve as early \emph{semantic anchors}, grounding the entire pipeline and ensuring that subsequent retrieval is guided by structured intent.

    \item \textbf{Keyword Extraction and Literature Retrieval.}  
    Based on the research plan, the Literature Agent extracts domain-specific concepts and generates an expanded keyword set consisting of explicit terms, latent concepts, and semantically related expressions derived through LLM-based inference. These terms are compiled into retrieval templates used by the ElasticSearch. Retrieval operates across multiple indexed paper fields with field-specific weighting: titles and abstracts receive higher relevance emphasis, while body text and authorship metadata broaden semantic coverage. This multi-field semantic retrieval mechanism significantly increases recall and precision. Additionally, the Literature Agent can optionally invoke the scientific network retrieval tool to expand the candidate set based on citation and reference relationships within the scientific network.

    \item \textbf{Relevance Ranking and Filtering.}  
    Retrieved documents are further evaluated by the Literature Agent using a multi-dimensional relevance and quality scoring mechanism. In addition to semantic relevance that captures topical alignment, methodological similarity, task-structure proximity, and latent domain adjacency, the agent also considers the scholarly quality of each paper. The evaluation incorporates factors such as citation impact, venue prestige, and empirical rigor. By integrating both conceptual relevance and scientific influence, the system prioritizes literature that is highly relevant and reliably impactful, ensuring that only strong and valuable papers progress to the next stage.

    \item \textbf{PDF Parsing and Key Point Extraction.}  
    Each retained paper undergoes structured PDF parsing to identify section headers, abstracts, methodological components and experimental results. The LLM extracts the paper's core contributions, innovations and empirical findings, refining these outputs through iterative self-verification. Cross-paper reasoning identifies conceptual variants, methodological connections, and experimental discrepancies, enabling a more integrated understanding of the literature landscape.

    \item \textbf{Draft Construction via Sketchboard Writing.}  
    Extracted insights are progressively incorporated into the Sketchboard, a structured drafting workspace managed by the Writing Agent. Through iterative refinement cycles, the agent develops coherent paragraphs, maintains narrative consistency, and enforces stylistic alignment across sections. The process transforms preliminary notes into a complete and analytically grounded literature review in Markdown format.
\end{itemize}

This workflow is supported by a decoupled multi-agent architecture composed of a Planning Agent, a Literature Agent, and a Writing Agent, coordinated by a top-level Agent Orchestrator responsible for dependency management, task scheduling, and global quality control. The modularity of this design allows for seamless integration of future specialized agents, such as mathematical reasoning or experimental analysis modules.


\begin{floatingfigure}[r]{0.6\textwidth}
  \centering
  \includegraphics[width=0.6\textwidth]{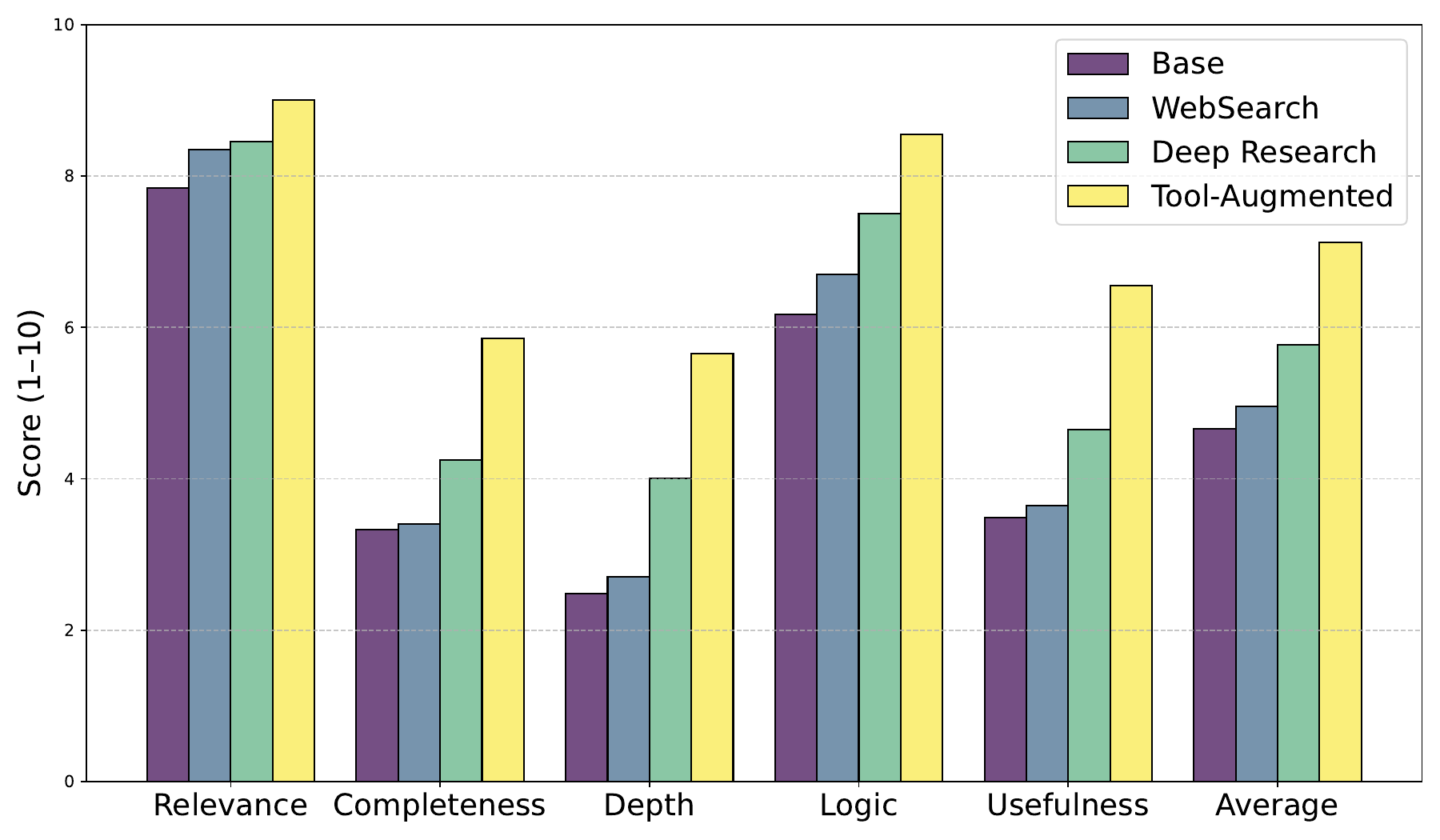} 
  \vspace{-4mm}
  \caption{Evaluation of Survey Quality Across Multiple Dimensions for Different Models}
  \label{fig:application}
\end{floatingfigure}

To exemplify the practical application value of the scientific network retrieval tool, we conducted a case study. This experiment, detailed in Figure \ref{fig:application}, evaluates the quality of surveys generated by four distinct methods: (1) a ``Base'' LLM (gpt-5) generating a survey from the query alone, without external retrieval; (2) a ``WebSearch'' enabled gpt-5, representative of the general-purpose systems we critiqued; (3) the ``Deep Research'' system, a state-of-the-art commercial agent (o4 mini DeepResearch) ; and (4) a ``Tool-Augmented'' gpt-5, where the base LLM was equipped with the retrieval results from the scientific network tool.
The generated surveys were evaluated by Gemini-2.5-pro across five dimensions: Relevance, Completeness, Depth, Logical Consistency, and Usefulness, with each dimension scored on a 1–10 scale.

The results compellingly demonstrate that retrieval quality is the primary determinant of survey quality. The ``Tool-Augmented'' model, representing perfect, relation-aware retrieval, achieves the highest scores across all dimensions, confirming this clear upper bound. The ``Deep Research'' system, while the strongest baseline, substantially outperforms the ``Base'' and ``WebSearch'' approaches, particularly in Completeness, Depth, and Logic.
Moreover, we observe a substantial performance gap between the ``Deep Research'' system and the ``Tool-Augmented'' model. This gap reinforces our central claim: even the most advanced commercial systems, which do not incorporate relation-aware retrieval, remain unable to reconstruct the full set of logically connected publications underlying a scientific topic. Their failure to capture these structural relations leads to surveys that are noticeably less complete and less coherent. By contrast, the scientific-network-augmented retrieval pipeline in our OmniScientist system is specifically designed to mitigate this deficiency by identifying a more comprehensive and structurally grounded set of relevant papers.

\subsection{Research Ideation}
The generation of novel research ideas plays a critical role in advancing scientific research. While recent advancements in LLMs have shown potential for this task, previous work in research ideation has suffered from several key limitations. Primarily, existing approaches have relied on simplistic methods, such as keyword co-occurrence or semantic similarity~\cite{sourati2023accelerating, tshitoyan2019unsupervised}. These techniques typically focus on identifying statistical associations within the literature or static concept representations, thereby overlooking the complex, contextual, and co-occurrence-based relationships that researchers construct. Some LLM-driven methods, on the other hand, propose and iteratively refine research ideas by leveraging relevant literature retrieved through techniques like semantic similarity~\cite{yang2024moose}. However, these methods fail to fully utilize the valuable network of scientific concepts, which reveals the nuanced relationships among literature arising from shared concepts. While some works do attempt to leverage this concept network~\cite{wang2024scimon, baek2025researchagent}, they are typically limited to retrieving only first-order neighbors directly related to the query. This narrow focus neglects deeper, higher-order relationships and more specific connections between concepts, which are essential for generating more comprehensive insights.

\begin{figure}[!ht]
\begin{center}
\includegraphics[width=14cm]{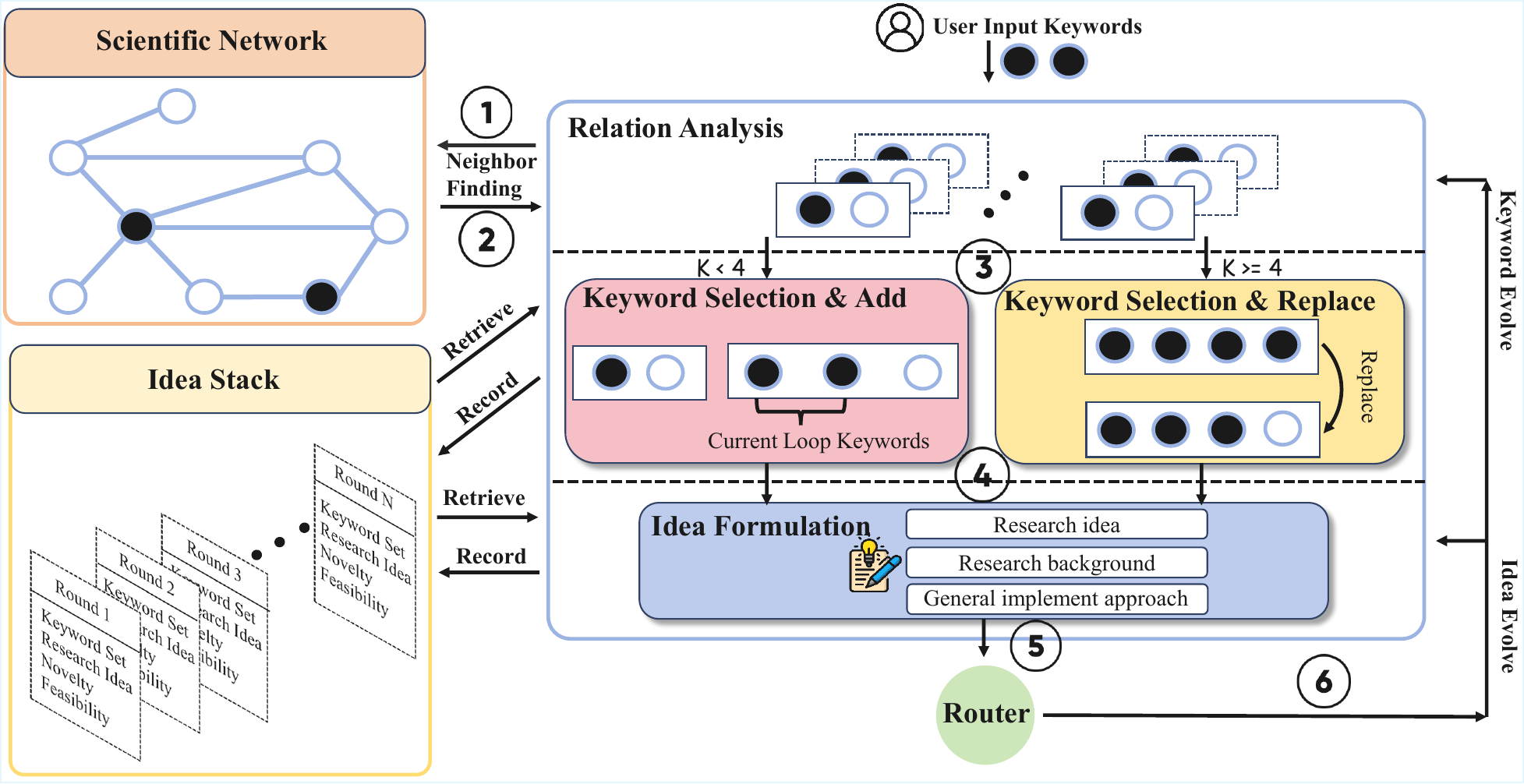}
\vspace{-6mm}
\caption{Overview of our Deep Ideation framework. In this figure, we set the maximum size of the keyword set to 4.}
\label{ideation1}
\end{center}
\end{figure}

In order to enable LLMs to effectively leverage the valuable network of scientific concepts, integrating their internal knowledge with human research achievements, we propose \textbf{Deep Ideation} framework. Within this framework, the LLM iteratively queries the scientific network through an explore-expand-evolve workflow, dynamically acquiring human knowledge embedded within the network. In parallel, the Idea Stack tracks the progression of ideas, offering an overarching perspective on the evolving research process, much like how human researchers refine their ideas over time through accumulated insights. The generated ideas are continuously refined through review feedback that aligns with the level of human expertise, ultimately resulting in a high-quality idea proposal. The overall process is illustrated in Figure~\ref{ideation1}.

\textbf{Key Components of Deep Ideation Framework}

There are four key components in the Deep Ideation Framework: the Scientific Network, the Relation Analysis Module, the Keyword Selection Module, and the Idea Formulation Module.
\begin{itemize}
    \item The Scientific Network is constructed based on the co-occurrence relationship of concepts in the literature. Here, we define concepts as the keywords found in the literature. Initially, the literature's title, abstract, and introduction are input into the LLM, and we prompt the LLM to extract relevant keywords from these sections. These keywords are then treated as nodes in the network. Subsequently, co-occurring keywords within the same literature are connected by edges.

    \item Relation Analysis Module is responsible for summarizing how co-occurring literature, through the process of human authorship, construct connections between keywords. Specifically, it analyzes the relationships between keywords and their neighboring terms as established in the literature, capturing the way these terms are linked in the context of scientific research.

    \item Keyword Selection Module plays a crucial role in steering the ideation process by selecting the most significant and impactful keywords to expand the initial set. Beyond merely refining the keyword collection, this module actively shapes the direction of the evolving idea, ensuring that it remains focused on the most promising avenues for both novelty and feasibility.

    \item Idea Formulation Module addresses a key gap in many existing approaches~\cite{sourati2023accelerating}, which often focus solely on keyword combinations without providing a complete, structured idea proposal. This module plays a critical role in synthesizing the selected keywords into a coherent and scientifically grounded idea proposal, transforming a set of keywords into a fully formed concept.
\end{itemize}
\textbf{Explore-Expand-Evolve Workflow}

\textbf{Explore:}
The process begins with an initial set of keywords \( K_0 = \{ k_1, k_2, \dots, k_n \} \), which are refined by identifying and analyzing their neighboring terms within the scientific network. To obtain the neighboring keywords, we define \( N(K_0) \) as the set of neighboring keywords for all \( k_i \in K_0 \). Since the number of neighbors for each keyword may be large, we limit the selection to the \( m \) neighboring terms, where \( m \) is a predefined maximum number. This gives us a set of neighboring keywords for each \( k_i \):

\[
N(K_0) = \{ N(k_1), N(k_2), \dots, N(k_n) \}
\]

Each \( N(k_i) \) is limited to the \( m \) neighbors. The \textbf{Relation Analysis Module} then analyzes the relationships between each pair of selected keywords \( (k_i, k_j) \) and their common co-occurrence across multiple papers. Given that multiple papers can share co-occurring keywords, the relationship \( R(k_i, k_j) \) between two keywords is derived by considering all the papers where both \( k_i \) and \( k_j \) appear, represented by \( \mathcal{P}(k_i, k_j) \):

\[
R(k_i, k_j) = g(k_i, k_j, \mathcal{P}(k_i, k_j))
\]

where \( \mathcal{P}(k_i, k_j) = \{ p_1, p_2, \dots, p_t \} \) represents the set of papers \( p \) that both \( k_i \) and \( k_j \) co-occur in, and \(g\) is a function that aggregate the relationship together.

\textbf{Expand:}
Following the exploration and relation analysis, the \textbf{Keyword Selection Module} is tasked with selecting the most significant keyword \( k_{\text{new}} \) to add to the current set \( K_t \), where \( k_{\text{new}} \in N(K_0) \). The selection is based on a comprehensive analysis of the relationship between the new keyword and the existing set of keywords. This new keyword is chosen by evaluating the relationships \( R(k_{\text{new}}, k_i) \) for each \( k_i \in K_t \), where the relationship between the newly selected keyword and an existing keyword is considered:

\[
R(k_{\text{new}}, k_i) = g(k_{\text{new}}, k_i, \mathcal{P}(k_{\text{new}}, k_i))
\]

The Keyword Selection Module outputs the selected keyword \( k_{\text{new}} \), the reason for the selection (based on its relationship to the current keyword set), and its connection to the existing keyword. The selected keyword is then added to the current keyword set:

\[
K_{t+1} = K_t \cup \{ k_{\text{new}} \}
\]

Subsequently, this updated set of keywords \( K_{t+1} \) and the \textbf{Idea Stack}, which contains all previous research iterations (including keyword sets and idea proposals), are input into the \textbf{Idea Formulation Module}. The Idea Formulation Module synthesizes the selected keywords into a coherent idea proposal, which includes the research background, research idea, and a general implementation approach. The idea proposal at time \( t \) is generated as:

\[
P_t = LLM(K_{t+1}, \text{prompt})
\]

where the prompt represent the prompt template for idea formulation module. The Idea Stack records each round’s progress, tracking keyword evolution, idea development, and evaluations, thus mirroring the iterative nature of human research.

\textbf{Evolve:}
The Evolve Mechanism triggers when the keyword set reaches a predefined length \( L_{\text{max}} \). At this point, the focus shifts to evolving  the keyword set or the idea proposal. The \textbf{Router} determines whether the focus should be on refining the keyword set or on adjusting the idea proposal. The Router decision is formalized as:

\[
\text{Next Action} = 
\begin{cases}
\text{Keywords Evolve} & \text{if } \text{Router} == \text{Evolve}(K_t) \\
\text{Idea Proposal Evolve} & \text{if } \text{Router} == \text{Evolve}(P_t)
\end{cases}
\]

During the evolution phase, the keywords in \( K_t \) are dynamically replaced based on insights from previous iterations. This evolution is represented by:

\[
K_{t+1} = (K_t \setminus \{ k_{\text{old}} \}) \cup \{ k_{\text{new}} \} \quad \text{or} \quad P_{t+1} = LLM(K_{t+1}, \text{prompt})
\]

The idea proposal is refined by incorporating new findings and emerging research trends, while the keyword set is updated iteratively to adapt to the evolving research context. This ensures that the generated ideas continue to evolve, progressively becoming more novel and feasible.

\textbf{Critic Model}

The Critic Model drives the iterative refinement of ideas in the Deep Ideation framework by providing expert-level evaluative feedback on generated proposals. This feedback loop ensures continuous improvement by aligning with domain-specific evaluation standards. Although LLMs can be used for reviews, they lack the nuanced, expert-level reasoning required for deep evaluation. To address this, we developed a 'Scientific Reasoning Simulation' prompt that enables the LLM to mimic the cognitive process of human reviewers, assessing novelty and feasibility based on existing research. This simulated reasoning is used to fine-tune the LLM, aligning its feedback with expert review standards.

Overall, we presented the Deep Ideation framework, which integrates LLMs with scientific networks to generate novel and scientifically grounded research ideas. By leveraging the relationships between keywords in scientific literature, our method ensures that generated ideas are both innovative and anchored in existing knowledge. The iterative workflow, enhanced by the Idea Stack, enables continuous idea refinement, mirroring the cognitive process of human researchers. Additionally, the critic model, trained on real-world feedback, provides critical evaluative input to ensure that the ideas are novel and feasible.

\subsection{Experiment Automation}

In the typical scientific research workflow, validating a novel idea necessitates rigorous experimentation against appropriate datasets and comparison with relevant baseline methods. While recent automated science platforms like FunSearch~\cite{romera2024mathematical} and AlphaEvolve~\cite{novikov2025alphaevolve} have excelled at optimizing solutions for specific problems, they often overlook this critical preceding step of resource selection. The few existing works dedicated to recommending datasets or baselines, such as DataFinder~\cite{viswanathan2023datafinder} and DataHunter~\cite{farber2021datahunter}, suffer from several key limitations. First, they typically rely solely on self-descriptions (e.g., the paper's abstract) for representation. However, the true academic positioning of a baseline or dataset is defined by both its self-description and, crucially, how other papers cite and describe it (its citation context). Second, these methods generally focus on recommending either baselines or datasets in isolation, ignoring the critical synergistic relationship between them. To address these gaps, OmniScientist proposes a novel two-stage framework for joint baseline and dataset recommendation. First, we construct a comprehensive representation for both baselines and datasets by extracting and combining their self-descriptions with their broader citation contexts. We then fine-tune an embedding model on this rich representation to perform coarse recall. To achieve fine-grained reranking, we explicitly model the interactions by extracting "paper-baseline-paper-dataset" citation chains. These paths are used to construct reasoning chains, upon which we fine-tune a Large Language Model (LLM) to generate an explainable final ranking.

\begin{figure}[!]
\begin{center}
\vspace{-6mm}
\includegraphics[width=14cm]{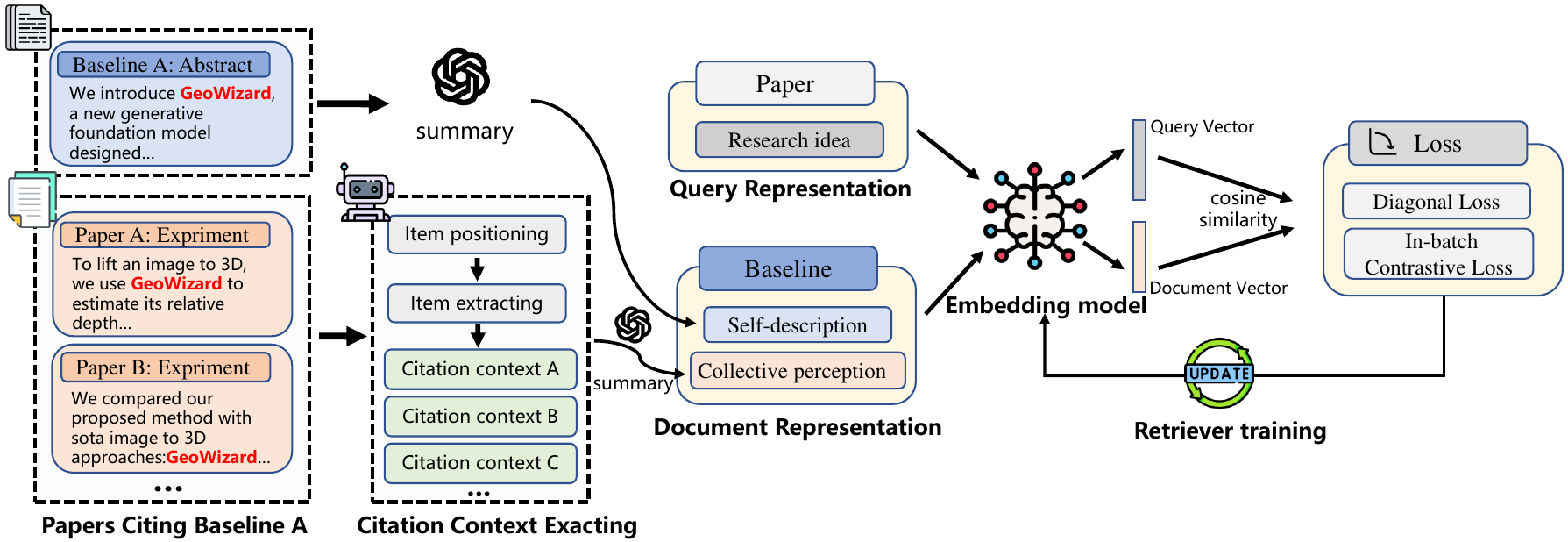}
\vspace{-4mm}
\caption{Illustration of Collective Perception Augmented Retrieval}
\label{experiment1}
\end{center}
\end{figure}

To implement the coarse recall stage, we construct a rich representation for each candidate (baseline or dataset) that moves beyond simple self-description, see Figure~\ref{experiment1}. We introduce a collective perception signal by first extracting all citation contexts for a given target from the experimental sections of papers in our corpus. Since the raw contexts can be numerous and noisy, we use a large language model to synthesize them into a concise summary. We then create the final target representation by concatenating its first-person self-description with this third-person collective perception. We finetune a bi-encoder retriever on these concatenated representations using a contrastive loss objective, training it to pull a query towards its true associated baselines and datasets.

\begin{floatingfigure}[r]{0.6\textwidth}
  \centering
  \includegraphics[width=0.6\textwidth]{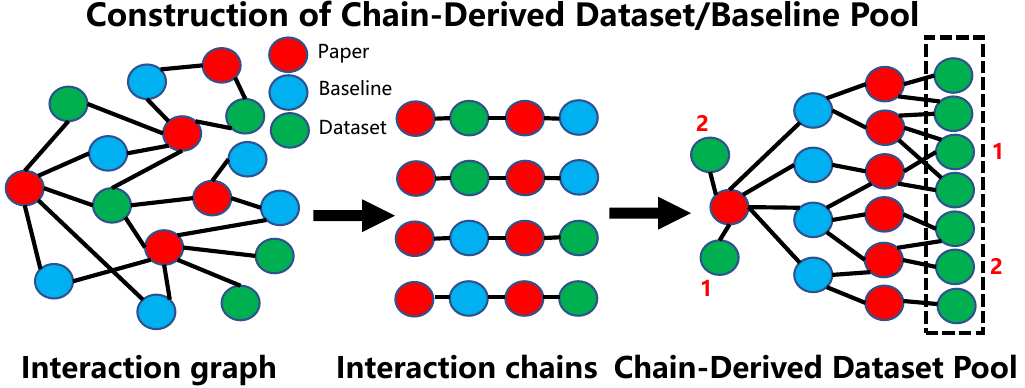} 
  \vspace{-4mm}
  \caption{Construction of Chain-Derived Dataset/Baseline Pool Analysis}
  \label{experiment2}
\end{floatingfigure}

In the fine-grained reranking stage, our objective is to leverage the synergistic relationship between baselines and datasets using a reasoning-augmented reranker, see Figure~\ref{experiment2}. For each candidate from the recall stage, we extract interaction chains from our scholarly knowledge graph. For example, to recommend a baseline $b$ for a query paper $p$, we find paths such as $\text{paper } (p) \to \text{dataset } (d) \to \text{paper}'($p'$) \to \text{baseline } (b)$. This chain explicitly connects the query paper to the candidate baseline via a shared dataset ($d$) that was also used by another paper ($p'$). We then finetune a large language model as a listwise reranker, training it to take the query, the candidate, and its evidential chains as input. The model's task is to generate an explicit reasoning chain that justifies the candidate's relevance, resulting in a final, interpretable, and precise ranking.

Once the appropriate datasets and baselines are identified, the system can proceed to the experiment execution phase. To facilitate efficient iterative optimization of the proposed idea, we constructed a multi-agent system for automated experimentation. This system comprises four specialized agents:
\begin{itemize}
    \item \textbf{Evolution Agent}: Responsible for generating new method code variants based on parent programs.
    \item \textbf{Sample Agent}: Constructs contextual prompts for the next iteration based on the evolutionary history and performance metrics.
    \item \textbf{Evaluation Agent}: Executes the code and measures various performance metrics.
    \item \textbf{Feedback Agent}: Analyzes execution errors or suboptimal performance and translates them into actionable suggestions for improvement.
\end{itemize}

\subsection{Scientific Writing}

\begin{figure}[h]
    \centering
    \includegraphics[width=0.9\linewidth]{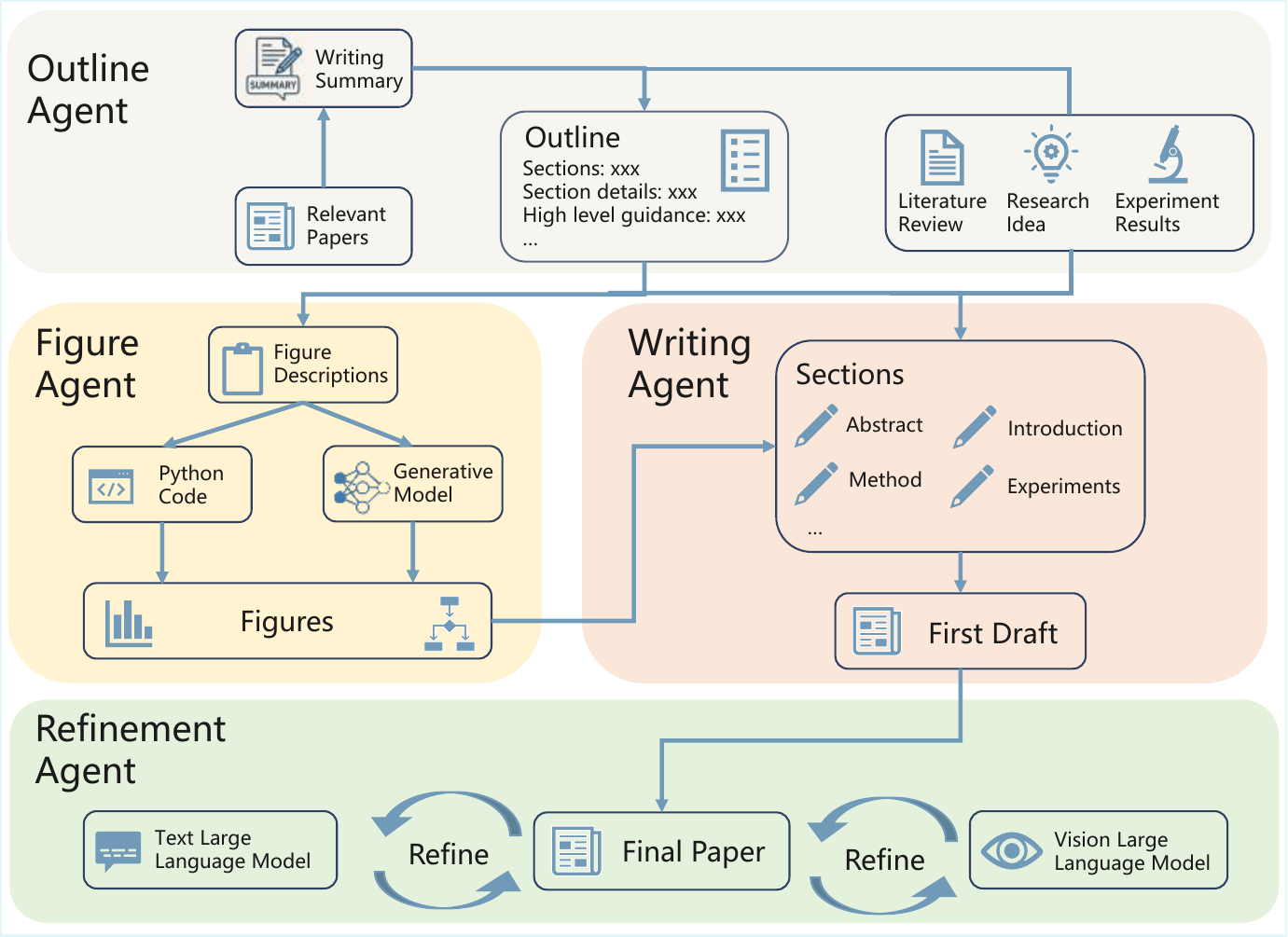}
    \caption{The overall framework of scientific writing}
    \label{fig:scientific_writing}
\end{figure}

\begin{figure*}[htbp]
    \centering

    \begin{subfigure}{0.48\textwidth}
        \centering
        \includegraphics[page=1,trim=90 60 90 60,clip,width=\linewidth]{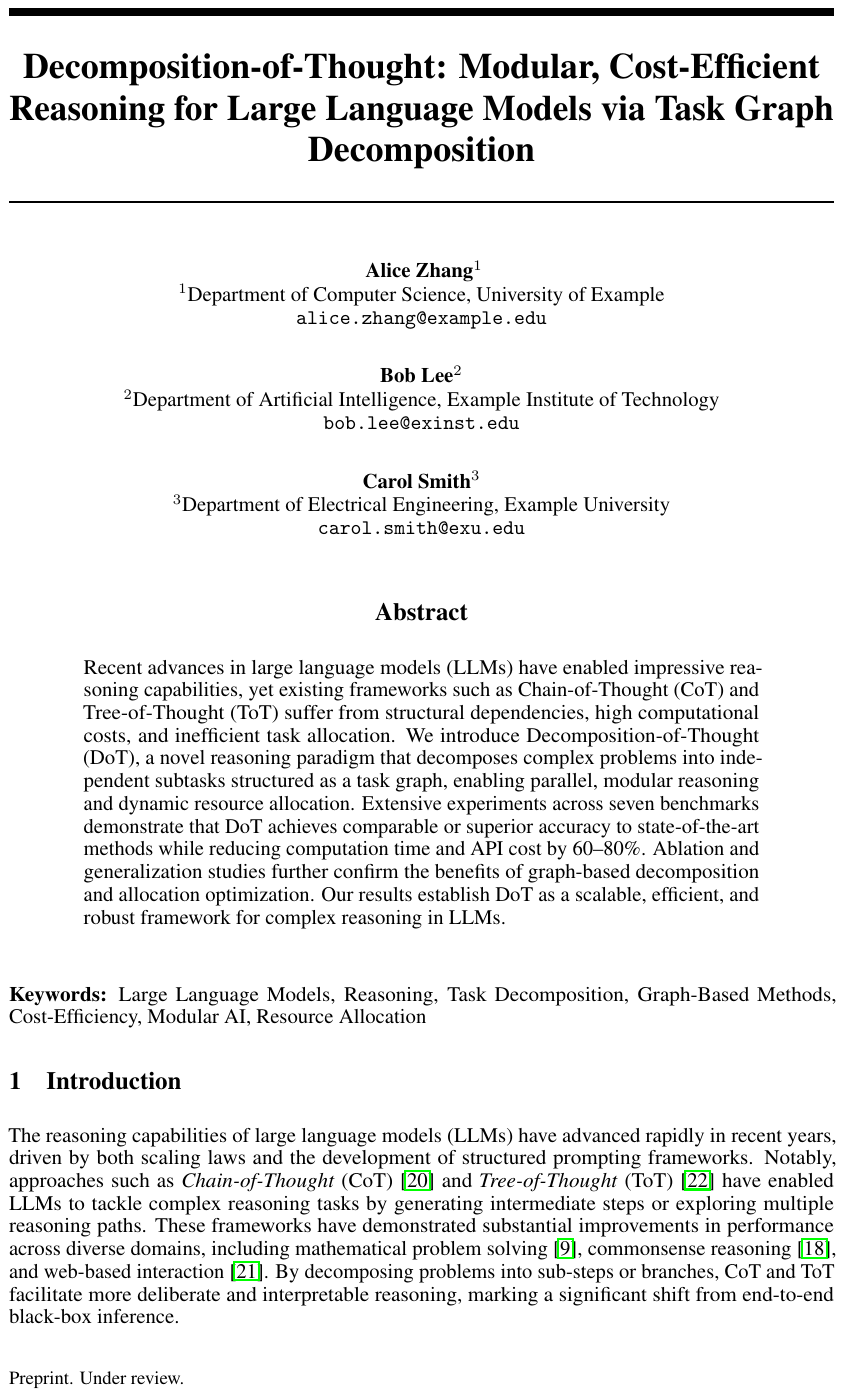}
        \caption{Page 1}
    \end{subfigure}
    \hfill
    \begin{subfigure}{0.48\textwidth}
        \centering
        \includegraphics[page=2,trim=90 60 90 60,clip,width=\linewidth]{Figures/scientific_writing_case.pdf}
        \caption{Page 2}
    \end{subfigure}

    \vspace{5mm}

    \begin{subfigure}{0.48\textwidth}
        \centering
        \includegraphics[page=8,trim=90 60 90 60,clip,width=\linewidth]{Figures/scientific_writing_case.pdf}
        \caption{Page 8}
    \end{subfigure}
    \hfill
    \begin{subfigure}{0.48\textwidth}
        \centering
        \includegraphics[page=10,trim=90 60 90 60,clip,width=\linewidth]{Figures/scientific_writing_case.pdf}
        \caption{Page 10}
    \end{subfigure}

    \caption{A Case of Scientific Writing}
    \label{fig:scientific_writing_case}
\end{figure*}

Previous works on AI scientists have developed a relatively mature pipeline for automated scientific paper writing. Typically, this process begins with either a human-defined or LLM-generated paper structure, followed by content generation and final refinement\cite{Lu2024TheAS,yamada2025ai,Tang2025AIResearcherAS,schmidgall2025agent}. However, this workflow presents two major limitations.
First, it \textit{fails to effectively learn from existing literature}, resulting in limited ability to emulate the linguistic style and structural conventions specific to a given subfield. Second, it  \textit{places insufficient emphasis on visual content}—while such systems can often generate data-related figures, they struggle to produce methodological or conceptual diagrams, which are crucial for clearly conveying research ideas.
These limitations arise because existing methods do not fully encode the established methodologies of human scientific writing into the AI scientist workflow.

To address the aforementioned issues, OmniScientist proposes a multi-agent framework that explicitly encodes two critical components of human scientific writing—the deep learning from related literature and the emphasis on high-quality visual communication (figures).
As shown in Figure \ref{fig:scientific_writing}, this framework consists of four synergistic agent subsystems: Outline Agent, Figure Agent, Writing Agent, and Refinement Agent.
To tackle the problem of insufficient learning from related works, the Outline Agent employs an LLM to summarize the most relevant papers, analyzing their writing strategies and structural patterns to derive a data-driven writing summary that informs the overall composition process.
To improve the visual quality of figures, the Figure Agent combines image generation models with Python-based script executors to produce methodology diagrams and data visualizations, respectively. Also, a Vision-Language Model (VLM) is employed to evaluate and iteratively refine the generated images, ensuring both accuracy and aesthetic coherence.
Together, these agents enable OmniScientist to autonomously generate coherent, visually enhanced, and field-adaptive research papers. The modular design promotes interpretability, extensibility, and domain transferability, making OmniScientist a powerful foundation for fully automated scientific writing.
The details of each agent are as follows:

\begin{itemize}
    \item \textbf{Outline Agent}:
    The Outline Agent analyzes the most relevant papers to learn their writing styles and structural patterns, generating an optimized writing summary. Based on the research idea, literature review, experiment results, and formatting requirements, it constructs a paper outline that defines section titles and details for each section. It also produces a high-level guidance specifying tone, style, and figure design preferences to guide the entire writing process.

    \item \textbf{Figure Agent}:
    The Figure Agent uses an LLM to generate a figure list from the outline, detailing each figure’s section, description, required data, and type (method or data). For method figures, it  then refines the textual description and invokes an image generation model; for data figures, it then writes and executes Python scripts to produce visualizations. Each figure is then verified by a Vision-Language Model (VLM) to ensure accuracy and visual quality.

   \item \textbf{Writing Agent}:
    The Writing Agent generates text section by section, utilizing the outline, generated figures, literature review, experiment results, and research ideas as input. If the author provides a bibliography (BIB) file alongside the literature review, the agent will use those existing references; otherwise, it will generate the BIB file based on the content of the literature review.  For each section, it first proposes a structure and then generates the content as a LaTeX-formatted output. The Agent subsequently evaluates the quality of the generated text and provides feedback; if the quality is deemed insufficient, the writing process is repeated.
    
    \item \textbf{Refinement Agent}:
    The Refinement Agent initiates a quality assessment, checking for length, format, and consistency issues, as well as its adherence to the provided guidance. Following this, it applies a single, comprehensive revision based on high-level feedback to the entire document. Afterward, it compiles the paper in LaTeX, automatically fixes compilation errors, and employs a VLM-based evaluation to assess figure quality and layout aesthetics, yielding a polished, publication-ready paper.

\end{itemize}

Overall, OmniScientist provides an intelligent multi-agent framework that successfully completes the automated scientific paper writing task. By integrating literature-informed structuring, high-quality figure generation, LLM-driven writing, and VLM-based refinement, it effectively encodes the complexity of the human scientific writing process and ensures the generated output meets standards of coherence and visual clarity. Crucially, this modular architecture facilitates seamless collaboration among specialized agents and promotes interoperability with both human researchers and other AI Scientist components, laying a critical foundation for building a robust AI research ecosystem. Figure ~\ref{fig:scientific_writing_case} shows some pages of a demo case created by our paper writing framework.

\subsection{Paper Review}
Recent advancements in Large Language Models (LLMs) have catalyzed the development of automated scholarly paper review (ASPR) systems\cite{lu2024ai, liang2024can, d2024marg, taechoyotin2025remor, zeng2025reviewrl, weng2024cycleresearcher, gao2024reviewer2, yu2024automated, zhuang2025large, hossain2025llms}, transitioning from general-purpose models to specialized agents. Current state-of-the-art approaches range from finetuned specialist models, such as OpenReviewer\cite{idahl2025openreviewer} and DeepReviewer\cite{zhu2025deepreview} designed to emulate expert tone, to hierarchical decomposition frameworks like TreeReview\cite{chang2025treereview} that enhance efficiency through structured questioning. Concurrently, sophisticated multi-agent systems, including MAMORX\cite{taechoyotin2024mamorx} and Agent Reviewers\cite{luagent}, simulate academic review committees with specialized, multimodal roles.

Despite this progress, significant gaps remain. Current systems, while incorporating external knowledge retrieval or agent-based discussion, often fail to provide fine-grained, verifiable traceability for their judgments. This creates a process that is difficult to inspect and undermines trust. Furthermore, existing iterative mechanisms are typically model-centric—employing reinforcement learning with judge-models or autonomous self-correction—rather than being designed for explicit human-in-the-loop (HITL) collaboration. Finally, while multimodal analysis has emerged, it frequently remains high-level, such as the ``Figure Critic'' role, and fails to detect subtle, fine-grained inconsistencies. Critically, these systems often ignore the real-world challenge of parsing artifacts, including OCR or Markdown errors, making them less reliable in practice.

\begin{figure}[!]
\begin{center}
\includegraphics[width=0.95\textwidth]{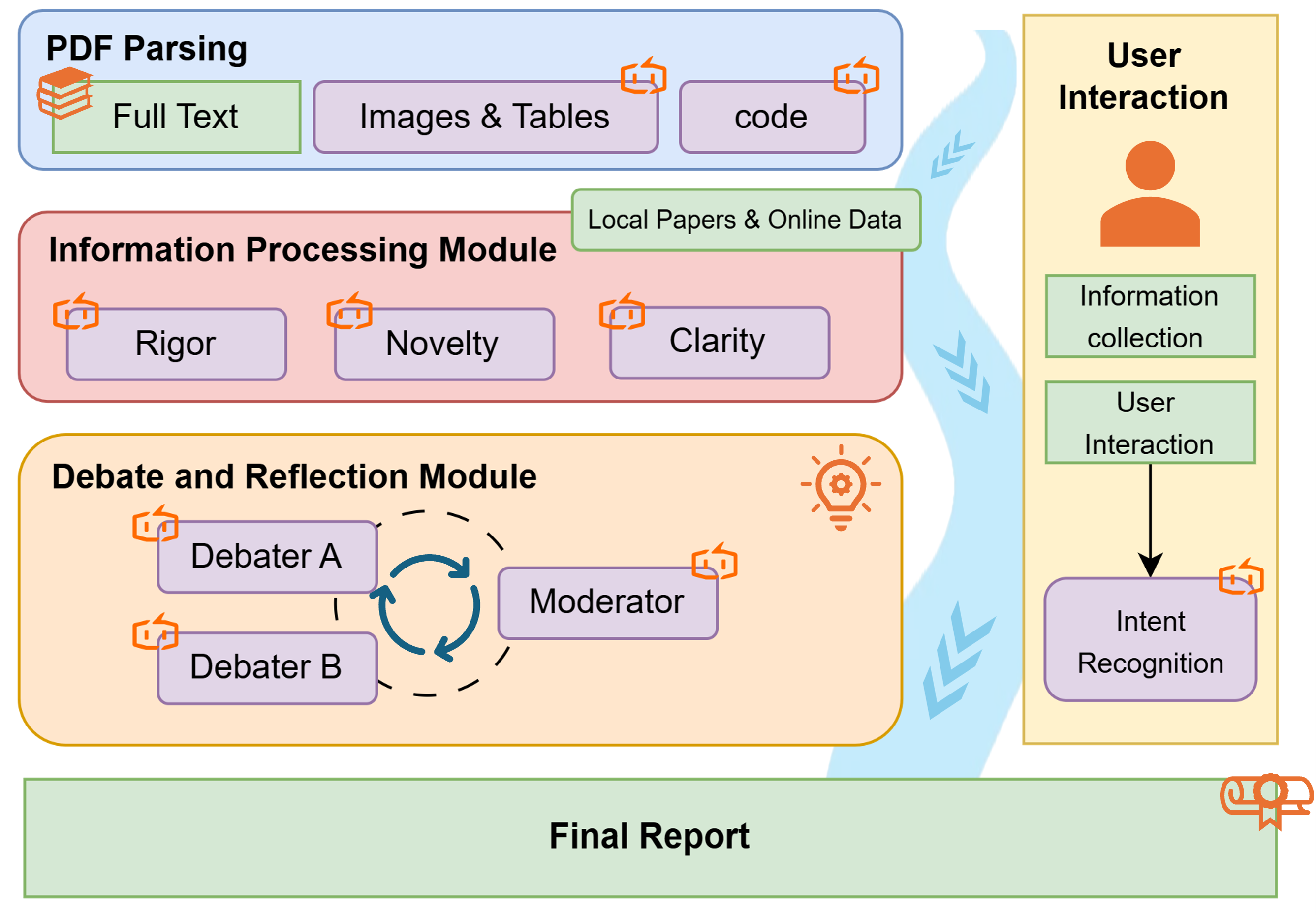}
\caption{Overview of \textbf{TIMAR} (Traceable and Interactive Multi-Agent Review) framework.}
\label{paper_review}
\end{center}
\end{figure}

To address these limitations, we propose a \textbf{Traceable and Interactive Multi-Agent Review (TIMAR)} system, based on the detailed design of the multi-agent paper review system presented. The overall system is illustrated in Figure~\ref{paper_review}. TIMAR implements academic peer review as a structured, evidence-driven, five-stage workflow.
The process begins with \textbf{(A) Input Pre-processing}, where the manuscript is parsed, and multimodal elements (figures and tables) undergo fine-grained consistency verification (e.g., cell-by-cell checks) to identify and tolerate parsing artifacts. Next, \textbf{(B) Retrieval Augmentation} builds a dedicated evidence pool by selecting key references from the paper and enabling both broad (Breadth) and targeted (Depth) retrieval from external source. This evidence informs \textbf{(C) Parallel Review}, where three distinct agents—\textbf{Novelty}, \textbf{Rigor}, and \textbf{Clarity}—independently generate initial, evidence-bound assessments. These drafts are then fed into the \textbf{(D) Merge and Iterative Revision} stage, which functions as a collaborative nexus. Here, an internal multi-agent debate and external Human-in-the-Loop (HITL) feedback (e.g., preference inputs) are treated as explicit instructions to refine the review. Finally, the \textbf{(E) Final Report and Annotation} stage synthesizes the converged text, automatically annotating key judgments with traceable citations $[n]$ that link directly to the manuscript or the retrieved evidence pool.

This architecture moves beyond simple automation to become a governed process, generating transparent and robust critiques for human collaboration. It directly tackles the core limitations we identified by delivering:
(1)Detailed and auditable fact-checking with full explainability;
(2)A novel framework for multi-round Human-in-the-Loop (HITL) interaction; and 
(3)
Reliable, in-depth analysis of multimodal elements, featuring explicit artifact tolerance.

\textbf{Verifiable Traceability and Explainability.}
The TIMAR framework moves beyond opaque judgments, achieving verifiable traceability through two core mechanisms. First, it applies an ``evidence-driven'' discipline to the parallel review agents (Novelty and Rigor). This discipline requires retrieval actions to gather external proof, preventing claims from being assessed in isolation. For example, it requires a ``Breadth $\rightarrow$ Depth'' retrieval sequence to verify strong "first-ever" claims against both broad academic fields and specific prior works.Second, the final output stage generates a traceable ``conclusion—evidence—citation'' chain. The system automatically marks key judgments in the review text (e.g., underlining) and links them with a numbered citation $...[n]$. This citation links directly to the evidence source, whether it is a specific location in the manuscript (e.g., ``Section 3, Table 2'') or a retrieved paper from the system's evidence pool.This two-pronged approach makes the agent's entire reasoning process transparent, allowing users to trace any conclusion back to its original evidence.

\textbf{Human-in-the-Loop Iterative Refinement.}
Instead of relying on autonomous self-correction, our system introduces an iterative refinement process explicitly designed for human-AI collaboration. This process is centered on the ``Merge and Iterative Revision'' stage, which functions as a central hub for collaboration.
This stage first merges the parallel reviews and generates an internal ``Debate.'' This debate, which produces ``optimistic'' and ``skeptical'' viewpoints, does not automatically resolve conflicts. Instead, its purpose is to highlight the most important or disputed points for a human operator (such as an Area Chair) to resolve.
This mechanism simulates the collaborative decision-making process inherent in human editorial committees, keeping the AI agent aligned with evolving community standards.
Crucially, the system is designed to accept ``User Preferences'' (e.g., questionnaires or structured feedback) as explicit ``revision instructions.'' This external guidance allows a human to steer the review's focus, tone, or depth in subsequent rounds. During this refinement loop, the system is prevented from performing new retrieval, ensuring that revisions remain grounded in the existing evidence base and focus on finalizing the review.

\textbf{Robust Multimodal and Artifact Analysis.}
To ensure high reliability, the framework performs a robust analysis of multimodal elements and structural artifacts before the main review begins. This task is handled by the "Input Pre-processing" stage, which performs fine-grained verification to build a trustworthy input.
This stage's methods include a ``first-image-then-text'' evaluation pipeline, establishing a visual baseline for figures before assessing their textual descriptions for consistency. It also uses a "cell-by-cell consistency" verification for tables, which cross-validates the semantic content of a table's image against its parsed text representation (e.g., HTML) to catch subtle data inconsistencies.
Most importantly, the entire system follows a strict ``artifact tolerance'' principle. This core rule requires that agents do not penalize paper quality for parsing failures, such as OCR errors, malformed formulas, or table misalignments. This crucial distinction enables the system to reliably operate on real-world documents and differentiate between author error and parser error.

\section{Building Co-evolution Systems of Human / AI scientists}

The preceding sections have detailed the individual functional modules of OmniScientist, ranging from literature review and ideation to automated experimentation. 
However, functioning in isolation, these components remain fragmented tools rather than a coherent intelligence. 
To transcend this fragmentation and construct a unified scientific ecosystem, effective mechanisms are required to orchestrate these diverse capabilities and bridge the gap between human and machine researchers.

In this section, we present the architectural backbone that connects these discrete modules. 
We first introduce the \textbf{Omni Scientific Protocol (OSP)}, which acts as the connective tissue for communication, collaboration, and credit attribution. 
Building on this foundation, we then detail the Closed-loop Multi-Agent System and Human-AI Collaboration Mechanism.
To validate the efficacy of the OSP, we present two comprehensive case studies: one demonstrating the system's capability for autonomous discovery in stochastic derivative estimation via the closed-loop workflow, and another highlighting the synergistic potential of human-AI collaboration in solving complex reasoning challenges.

\subsection{Protocol}
\label{sec:protocol}

For a complex and large-scale AI Scientist Ecosystem that integrates numerous functional modules, a protocol capable of connecting all components and integrating all capabilities is of paramount importance. The general-purpose \textit{Model Context Protocol (MCP)}~\cite{anthropic_mcp2025} provides interfaces for agent-to-tool interactions, while the \textit{Agent-to-Agent (A2A)}~\cite{A2A_Protocol_2025} protocol establishes the foundation for inter-agent communication. 
Together, these provide the core infrastructure for scientific research protocols.
Building upon them, the \textit{Scientific Context Protocol (SCP)}~\cite{scp2025} has been proposed to provide a standardized research workflow through a centralized \textit{SCP Hub}.
This enables efficient coordination between science-oriented applications and external research assets such as laboratory instruments, databases, LLMs, and specialized computational models.

However, scientific research is not merely a data transmission workflow. It is a reasoning-aware coordination process, deeply interwoven with human intuition, collaborative debate, rigorous provenance tracking, and intellectual credit attribution.
A truly effective protocol must enable diverse agents, tools, data sources, and even human researchers to \textbf{co-evolve within a unified scientific context}, thereby supporting the systemic research capabilities envisioned for an AI Scientist Ecosystem.
Yet, under this setting, existing protocols reveal four fundamental deficiencies:

\begin{enumerate}
    \item \textbf{Human-is-External:} Current protocols treat human scientists as \emph{users} or \emph{operators} external to the system, rather than as the \emph{highest-level cognitive agents} within it. Human-AI interactions remain fragmented and non-protocolized, occurring through ad-hoc interruptions, overrides, or external interventions.

    \item \textbf{Collaboration-is-Dark:} Essential collaborative activities, such as group discussions, peer reviews, and mentorship, take place outside the protocol layer (e.g., via Slack, WeChat, or in-person meetings). Consequently, the most critical phase of consensus formation remains opaque and untraceable, breaking the provenance chain of scientific reasoning at its very origin.

    \item \textbf{Credit-is-Ambiguous:} Current protocols focus solely on whether a task is completed, ignoring who contributed intellectually. While data provenance can answer ``where did this data come from?'' it fails to answer ``where did this discovery originate?'' or ``who deserves credit for this insight?''.
\end{enumerate}

To address these foundational issues, we propose the \textbf{Omni Scientific Protocol (OSP)}, a novel protocol framework specifically designed for scientific research scenarios. OSP unifies \textbf{Human-AI collaboration (H-AI)} and \textbf{intellectual credit attribution}, enabling transparent, accountable, and semantically grounded coordination across the entire scientific workflow.

\subsubsection{From External Users to Internal Participants}

OSP fundamentally redefines the role of the human scientist at the protocol level. Instead of being treated as an external operator, the human is positioned as an internal participant: the highest-level decision-making entity within the ecosystem.

To enable this, OSP introduces a Unified Participant Model. In this model, the protocol no longer distinguishes between ``AI'' and ``human'' entities. Both are abstracted as a common type called Participant. Human scientists (\texttt{Human\_Participant}) and AI scientists (\texttt{AI\_Scientist\_Participant}) hold equal protocol-level status and can both send and receive messages symmetrically within the same communication fabric.

This design marks a fundamental shift in the nature of human–AI interaction. Communication is no longer hierarchical, where humans issue commands through interfaces, but peer to peer, where AI agents can proactively initiate asynchronous negotiations with human participants. As a result, human intuition, judgment, and decision making are no longer opaque operations external to the system. They become integral, auditable, and traceable components within the protocol itself.

To support this new form of asynchronous, long-horizon human–AI interaction, we designed a set of protocol performatives to replace the simple \texttt{request} and \texttt{inform} actions in traditional protocols. These performatives are specifically designed to capture the critical human–AI negotiation events that occur in scientific research activities:

\begin{itemize}
    \item \textbf{\texttt{REQUEST\_REVIEW(artifact\_id, criteria)}}:  
    The AI agent proactively sends this request to one or more \texttt{Human\_Participant} entities, asking for a review of a critical scientific artifact, such as a draft survey paper or a piece of algorithmic code. Upon sending this request, the agent's task state transitions to \texttt{WAITING\_FOR\_HUMAN}, awaiting expert feedback.

    \item \textbf{\texttt{REQUEST\_DECISION(task\_id, options)}}:  
    When an AI agent encounters a crucial branching point during exploration (for example, discovering two distinct but potentially feasible synthesis pathways), it can use this performative to request a high-level decision from a \texttt{Human\_Participant}, providing the relevant information (\texttt{options}) in a structured format.

    \item \textbf{\texttt{APPROVE(artifact\_id, version\_hash)}}:  
    A standard approval receipt returned by a \texttt{Human\_Participant}. This protocol message, including the version hash, is permanently recorded in the provenance chain and serves as a formal validation for subsequent steps.

    \item \textbf{\texttt{REJECT(artifact\_id, reason)}}:  
    A standard rejection receipt returned by a \texttt{Human\_Participant}. The \texttt{reason} field, which can be structured data or natural language, is itself valuable scientific information and can trigger the AI agent to reflect, re-plan, or initiate a new round of exploration.
\end{itemize}

In this way, human \emph{approval} or \emph{rejection} is no longer merely a click in a user interface, but a referable, legally significant protocol event that guarantees full-chain traceability in scientific workflows.

\subsubsection{A Centralized Hub Enabling Multi-participant Engagement}

After redefining the notion of a \textit{participant}, OSP introduces a \textbf{centralized Hub as the foundational infrastructure of the protocol} to address the challenge of multi-party scientific collaboration. In real-world research, collaboration is rarely a one-to-one, linear process. Instead, it inherently exhibits a complex \textbf{many-to-many (M-to-N)} structure. A single research project (\texttt{Project}) often involves multiple human scientists (e.g., PIs, PhD students, collaborators) and a variety of AI agents (e.g., data analysis agents, literature review agents). Existing peer-to-peer or bus-style agent communication protocols lack mechanisms to effectively manage such multi-party scientific teams.

To this end, the \texttt{Hub} is not merely a message broker; it serves three critical roles:

\begin{enumerate}
    \item \textbf{Identity and Project Registry:}  
    The \texttt{Hub} manages the unified registration of both \texttt{Human\_Participant} (human scientists) and \texttt{AI\_Agent\_Participant} (AI agents). More importantly, it registers and maintains the definition of each research \texttt{Project}. This design is essential, as it establishes clear participant boundaries and scopes of work for every research activity. Without explicit project demarcation, attribution of contributions would be impossible.

    \item \textbf{Message Exchange and Distribution Center:}  
    The \texttt{Hub} functions as the central node for the exchange, routing, and archival of all protocol-level communications. Within OSP, any participant may transmit a protocol message (e.g., \texttt{REQUEST\_REVIEW}) directly to the \texttt{Hub}. The Hub then routes this message, based on the associated project context, to one or more appropriate recipients. This star-shaped topology transforms a conventional $N \times N$ communication mesh into a scalable and manageable $N \times 1$ model, dramatically improving both extensibility and robustness.

    \item \textbf{Immutable Process Recorder:}  
    Because all human-AI, AI-AI, and even human-human interactions (as the protocol expands) must be mediated through the \texttt{Hub}, it naturally becomes the \textbf{single source of truth} for the entire scientific workflow. 
    It enforces the recording of every critical step, decision, and cognitive action, thereby providing what we term \textbf{forced auditability}, the fundamental technical guarantee for contribution provenance and accountability.
\end{enumerate}

In summary, the introduction of the \texttt{Hub} architecture transforms research activities from isolated one-to-one interactions into a manageable many-to-many collaboration framework. It is not merely a communication backbone but the anchor of identity, project, and process integrity, forming the foundational layer upon which human–AI collaboration and contribution tracing can be built.

\subsubsection{From Data Provenance to Contribution Provenance}

Building upon the collaborative framework unified and managed by the \texttt{Hub}, OSP establishes a complete mechanism for the transition \textbf{from data provenance to contribution provenance}. After all, any scientific platform that fails to clearly define intellectual contributions can hardly gain the trust and adoption of real-world researchers.

Unlike traditional protocols, OSP does not rely on a single, superficial log file to determine contribution. Instead, it leverages the project management core---the \texttt{Hub}---to construct and maintain a unified \textbf{Long Scientific Context}.

Since every activity, discussion, and procedural execution by all participants (both human and AI) must go through the \texttt{Hub}, it can capture and record every aspect of a project's lifecycle. This context encompasses not only the final research outputs but, more importantly, all intermediate data and related resources generated throughout the scientific process, such as cited literature, executed experiments, generated logs, charts, written code, and even discussion records among team members.

To enable contribution attribution, we define the \texttt{ScholarlyObject} as the fundamental carrier within the protocol. It represents the smallest unit of intellectual value within a scientific activity (for example, a \texttt{Hypothesis}, \texttt{CodeBlock}, or \texttt{Artifact}) and serves as the container of contribution.
The key mechanism that enables provenance tracking is that every \texttt{ScholarlyObject} must carry an immutable \texttt{ContributionLedger}. This ledger is a chronological record of intellectual actions, documenting each \texttt{Participant} (human or AI) who performed an action (such as \texttt{create}, \texttt{refine}, \texttt{propose}, or \texttt{approve}) along with the corresponding timestamp.

A simplified example of a \texttt{ContributionLedger} is shown below, illustrating the evolution of a hypothesis:

\definecolor{BlueViolet}{RGB}{138,43,226}  

\begin{tcolorbox}[notitle, sharp corners, breakable, 
     colframe=BlueViolet, colback=white, 
     boxrule=3pt, boxsep=0.5pt, enhanced, 
     shadow={3pt}{-3pt}{0pt}{opacity=0.3},
     title={},]
     \footnotesize
     {\fontfamily{zi4}\selectfont
     \spaceskip=0pt plus 0pt minus 0pt 
     
\begin{lstlisting}[breaklines=true,showstringspaces=false]
"ContributionLedger": [
  {
    "participant_id": "Human_A_ID (PhD Student Bieber)",
    "action": "PROPOSE_HYPOTHESIS",
    "timestamp": "..."
  },
  {
    "participant_id": "AI_Reviewer_ID (Review Agent)",
    "action": "refine_statement",
    "timestamp": "..."
  },
  {
    "participant_id": "Human_B_ID (Advisor Frank)",
    "action": "APPROVE",
    "timestamp": "..."
  }
]
\end{lstlisting}
}
\end{tcolorbox}

By binding each \texttt{ScholarlyObject} to its \texttt{ContributionLedger}, OSP establishes an unbroken \textbf{chain of contribution}. When a downstream agent is assigned to verify a given hypothesis, it is protocol-enforced to reference the original object and its complete ledger. This ensures that, regardless of how automated the subsequent research process becomes, the final scientific result (\texttt{Result}) can always be transparently traced back to all contributors---such as \textit{Human\_A}, \textit{AI\_Reviewer}, and \textit{Human\_B}.

\subsection{Closed-loop Multi Agent System}

\begin{figure}[!ht]
\begin{center}
\vspace{-3mm}
\includegraphics[width=14cm]{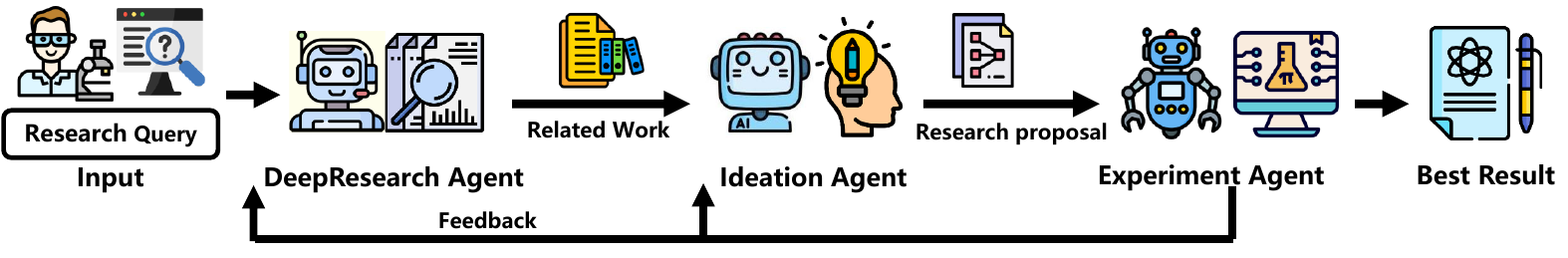}
\vspace{-6mm}
\caption{Overview of Closed-loop Multi Agent System}
\label{closed-loop1}
\end{center}
\end{figure}

Existing work in the AI Scientist domain, such as FunSearch~\cite{romera2024mathematical} and AlphaEvolve~\cite{novikov2025alphaevolve}, has primarily focused on the deep, evolutionary refinement of algorithms. These systems typically construct an evolutionary framework to iteratively improve a target algorithm, effectively framing scientific discovery as a standalone search and optimization problem. However, this isolationist paradigm overlooks the extensive collaborative mechanisms and supporting infrastructures, such as related papers, that constitute the foundation of human research. Consequently, existing AI scientists operate as solitary entities that fail to form a cohesive scientific ecosystem. By relying heavily on internal, pre-existing knowledge and ignoring the collective intelligence embedded in the scientific community, these systems are susceptible to converging on local optima, missing opportunities for more transformative, knowledge-grounded innovations. To address this gap, we propose a \textbf{Closed-loop Multi-Agent System} that integrates the specialized capabilities of DeepResearch, Ideation, and Automated Experimentation, see Figure~\ref{closed-loop1}. This framework is designed to balance deep algorithmic evolution with broad, knowledge-driven exploration, ensuring that innovation is both grounded in existing science and empirically validated.

This integrated system operates as a synergistic collective of specialized agents, each managing a critical phase of the scientific discovery lifecycle. The workflow commences with the \textbf{DeepResearch Agent}. Given an initial research topic or problem, this agent leverages our curated scientific network and advanced literature review pipeline to conduct a comprehensive survey. It synthesizes the current state-of-the-art, identifies established methodologies, and pinpoints unresolved challenges or gaps in the literature. This synthesized knowledge is then passed to the \textbf{Ideation Agent}, which utilizes the Deep Ideation framework to explore the scientific concept network, generating novel research hypotheses that are explicitly grounded in the context of prior work. Once a high-potential idea is formulated, the \textbf{Experiment Agent} takes responsibility. As detailed in our Experiment Design module, this agent executes the necessary experiments to validate or invalidate the hypothesis. Crucially, this process is not linear but a \textbf{closed loop}. The empirical results, error logs, and performance data generated by the Experiment Agent are fed back into the system. This new information serves as critical input for the Ideation Agent to refine or discard the hypothesis, or for the DeepResearch Agent to initiate a new, more targeted literature search to understand anomalous results. This iterative cycle of research, ideation, and experimentation allows the system to autonomously navigate the scientific landscape, progressively refining its understanding and converging on genuinely novel discoveries.

\subsection{Case Study: Variance Reduction in STDE via Closed-Loop Experiment}

To demonstrate the practical efficacy and distinct advantages of our closed-loop multi-agent system, we conducted a case study targeting the improvement of a highly influential, state-of-the-art method: the \textbf{Stochastic Taylor Derivative Estimator (STDE)}~\cite{shi2024stochastic}. This method, recognized as a best paper at NIPS 2024, provides an efficient amortization technique for arbitrary differential operators. While powerful, the accuracy of STDE is fundamentally reliant on standard Monte Carlo (MC) sampling to estimate complex expectations. The well-known limitation of MC sampling is its probabilistic convergence rate of $O(1/\sqrt{N})$, which can introduce significant variance and thus limit the precision of the solution, especially in high-dimensional settings. Our objective was to leverage our AI Scientist frameworks to discover and implement a principled modification that significantly reduces this estimation error.

\begin{floatingfigure}[r]{0.6\textwidth}
  \centering
  \includegraphics[width=0.6\textwidth]{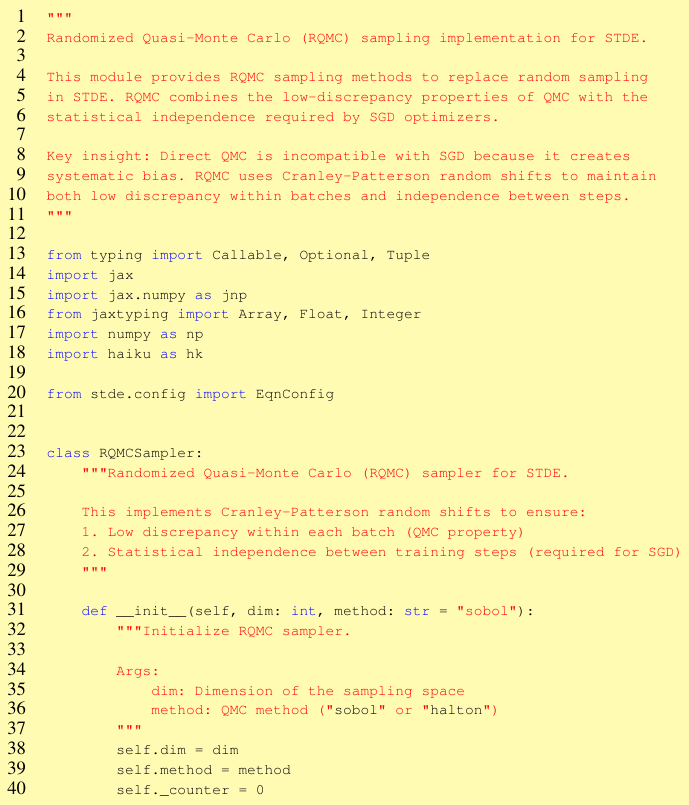} 
  \vspace{-4mm}
  \caption{Code of Quasi-Monte Carlo Generated by OmniScientist}
  \label{experiment2}
\end{floatingfigure}

We first deployed AlphaEvolve, which represents the paradigm of internal algorithmic evolution. It diligently explored variations of the existing STDE architecture, focusing on optimizing hyperparameters, testing different neural network architectures for the score function, and experimenting with alternative optimizers. In parallel, we deployed our \textbf{OmniScientist} system. Its DeepResearch Agent began by conducting a broad literature survey on "variance reduction," "stochastic derivative estimation," and "high-dimensional integration." The system identified a powerful, long-standing field of research, Quasi-Monte Carlo (QMC) methods, that was not utilized in the original STDE paper. The Ideation Agent hypothesized that the core MC sampler in STDE could be directly replaced with a QMC sampler. QMC methods utilize low-discrepancy sequences that cover the sample space more uniformly than the pseudo-random numbers of standard MC. This deterministic design leads to a superior theoretical convergence rate, often approaching $O(\log(N)^k / N)$, which translates directly to lower estimation error for the same computational budget.

\begin{wraptable}{r}{0.68\textwidth}
    \centering
    \caption{Error Comparison for AllenCahnTwobody equation}
    \label{tab:bs16}
    \begin{tabular}{@{}lcccc@{}}
    \toprule
    Method    & 100 D       & 1000 D      & 10000 D     & 100000 D    \\ \midrule
    STDE  & 0.008730 & 0.002620 & 0.003440 & 0.002500 \\
    Alphaevolve & 0.007859 & 0.001654 & 0.002059 & 0.003041\\
    Omniscientist & \textbf{0.006780} & \textbf{0.000579} & \textbf{0.000572} & \textbf{0.001210} \\
    \bottomrule
    \end{tabular}
\end{wraptable}

The experimental results, summarized in Table~\ref{tab:bs16} for the AllenCahnTwobody equation, validate this knowledge-driven approach. The AlphaEvolve variant, despite its extensive search, yielded only a marginal improvement over the STDE baseline. It successfully fine-tuned the existing method but remained confined within its original conceptual boundaries, thus failing to address the fundamental bottleneck of MC variance. In stark contrast, the OmniScientist-proposed solution, which introduced the external concept of Quasi-Monte Carlo sampling, achieved a dramatic and consistent reduction in solution error across all tested dimensions. This case study demonstrates the superior capability of OmniScientist to achieve significant scientific breakthroughs by actively seeking, integrating, and applying external knowledge from the broader scientific literature.

\subsection{Human-AI Collaboration}

Human expertise remains indispensable in the scientific discovery process, even as AI scientist systems continue to advance. Human scientists possess domain knowledge, methodological intuition, and practical experience that allow them to recognize blind spots, detect conceptual drift, and identify low-value or erroneous reasoning patterns that AI systems may overlook. Incorporating human insights therefore not only compensates for the limitations of automated reasoning but also stimulates new ideas and alternative perspectives, ultimately enhancing the scientific workflow.

Existing AI scientist ecosystems provide only limited support for such collaboration. In systems such as CRISPR-GPT~\cite{qu2025crispr}, human feedback is allowed through a lightweight user-proxy that accepts natural language instructions, yet the underlying interaction protocol remains coarse and under-specified. Other systems, such as Virtual Lab~\cite{swanson2025virtual}, introduce more explicit points of human involvement, including goal setting and review of intermediate outcomes. However, these workflows largely position humans as external supervisors rather than integrated team members, and their rigidly predefined interaction patterns restrict flexibility during complex scientific exploration.

Building on the unified protocol described in Section~\ref{sec:protocol}, we propose a more systematic and adaptable approach to human-AI collaboration. Our design seeks to support fine-grained human intervention while preserving the autonomy and exploratory capabilities of AI scientists. Specifically, our system accommodates the following key scenarios:

\begin{itemize}

\item \textbf{Multi-human participation.}
Many existing systems assume a single human paired with a multi-agent AI team, which limits collaborative potential. In real scientific practice, however, creative progress often emerges from the complementary expertise of multiple human researchers. Our system explicitly supports multi-human interaction, enabling experts from different fields to contribute diverse perspectives and engage collaboratively with the AI team.

\item \textbf{Human validation of critical outputs.}
When the AI scientist produces consequential artifacts such as draft papers, algorithmic designs, or experimental conclusions, it can request human review. Such feedback is treated not as a simple annotation but as part of the scientific record that shapes subsequent planning, refinement, or redirection.

\item \textbf{Human involvement in decision points.}
During exploration, the AI scientist may encounter branching pathways or ambiguous solution strategies. In these cases, it can query human experts for guidance. Human judgments rooted in experience, intuition, or theoretical understanding help steer the system toward meaningful and scientifically coherent trajectories.

\item \textbf{Formal approval and rejection.}
Human scientists may explicitly approve or reject AI-generated outcomes. These decisions carry procedural weight: a rejection triggers structured reflection, re-planning, or re-execution, and both the decision and its rationale are recorded in the research context to ensure transparency and reproducibility.

\end{itemize}

Through these mechanisms, our system supports a rich spectrum of collaboration patterns ranging from low-level iterative discussions to high-level review and strategic decision-making. The result is a flexible, protocol-driven framework that integrates human expertise with autonomous AI reasoning, better aligning AI scientist ecosystems with the complex, dynamic, and inherently collaborative nature of real scientific research.

\subsection{Case Study: HLE Challenge via Human-AI Collaboration}

To investigate the potential of human-AI collaboration in solving complex scientific research problems, we designed and conducted a delicate case study based on the \textbf{Humanity's Last Exam (HLE)\cite{phan2025humanity}}. Specifically, we constructed three controlled experimental conditions corresponding to three task-solving modes: (i) \textbf{AI Solo Mode}, in which the model solves the questions independently, (ii) \textbf{Human Solo Mode}, in which human participants solve questions on their own, and (iii) \textbf{Human-AI Collaboration Mode}, in which humans and an LLM jointly solve questions through structured interaction.
The Human-AI Collaboration Mode was designed as an interactive setting inspired by Tree of Thoughts (ToT)\cite{yao2023tree}. Participants engaged in multi-round interactions with the model. In each round, the LLM generated three reasoning paths or intermediate results for the participant to evaluate. Participants could choose among these paths or provide feedback to guide further refinement. The model then produced three new reasoning paths in the next round. This process continued until the participant submitted the final answer. The system automatically recorded the entire interaction process, response time, and answer correctness.

The study recruited 10 PhD-level participants. Each participant completed 10 HLE questions spanning computer science and artificial intelligence. For every individual, five questions were completed in Human Solo Mode and five in Human-AI Collaboration Mode. Question assignment followed a cyclic matrix design, ensuring that each question was answered by five participants in the Solo condition and by another five in the Collaboration condition. This allocation guaranteed balance from both the question and participant perspectives and enabled systematic comparison between individual and collaborative performance. The 10 HLE questions have been carefully selected from the computer science/AI category, ensuring that they cover various sub-fields such as machine learning, spatial recognition, database and query processing, algorithms and data structures, etc. in order to fully assess the reasoning potential of the three modes across different domains. For Human-AI Collaboration Mode and AI Solo Mode, we choose GPT-5~\cite{OpenAI2025GPT5} as the LLM to process questions independently or collaboratively with human participants. 

\begin{floatingfigure}[r]{0.5\textwidth}
  \centering
  \includegraphics[width=0.5\textwidth]{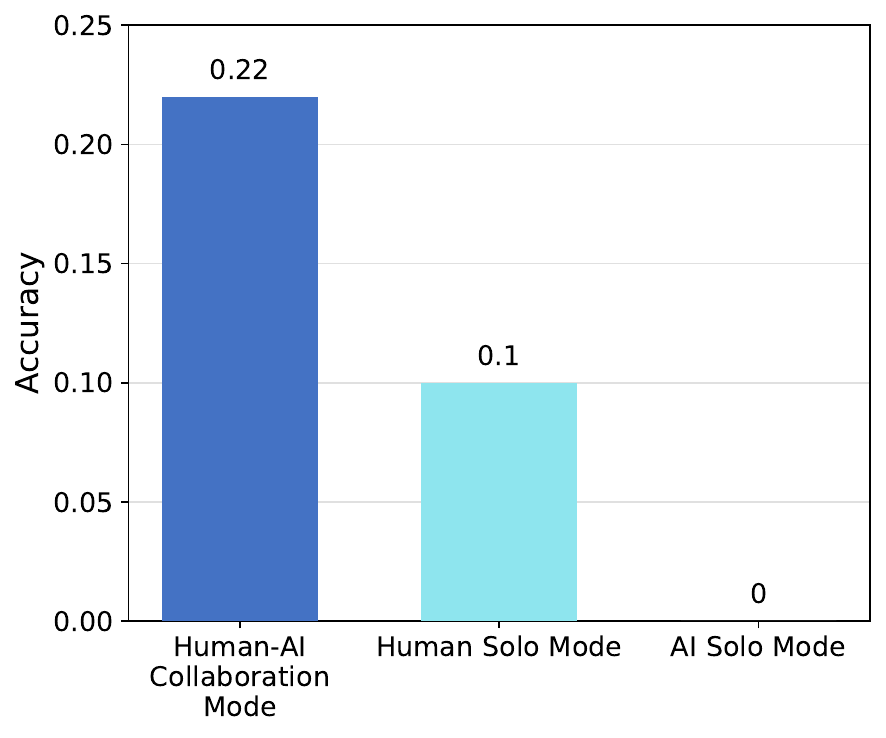} 
  \vspace{-4mm}
  \caption{HLE average accuracy across modes}
  \label{fig:hle_accuracy}
\end{floatingfigure}

The experimental platform provided a unified frontend interface through which human participants interacted with the system. All interaction data and results were automatically recorded and stored as timestamped JSON files for subsequent analysis and evaluation. This setup enabled fine-grained control over the collaborative process and offered a structured data foundation for examining the dynamic mechanisms of synergic human-AI collaboration in complex tasks.



Above in the picture~\ref{fig:hle_accuracy} are the result of our case study. The results demonstrate that human-AI collaboration significantly outperforms solo human efforts, which highlights the effectiveness of integrating human expertise with AI capabilities, particularly in complex scientific tasks. The collaborative mode's higher accuracy suggests that iterative human-AI interactions, as implemented in the study, effectively leverage the strengths of both parties. In contrast, the LLM mode's consistent accuracy of 0.0 implies the limitations of AI when operating independently, emphasizing the need for human guidance to enhance AI performance in nuanced scientific contexts.

\begin{figure}[!h]
    \centering
    \includegraphics[width=\textwidth]{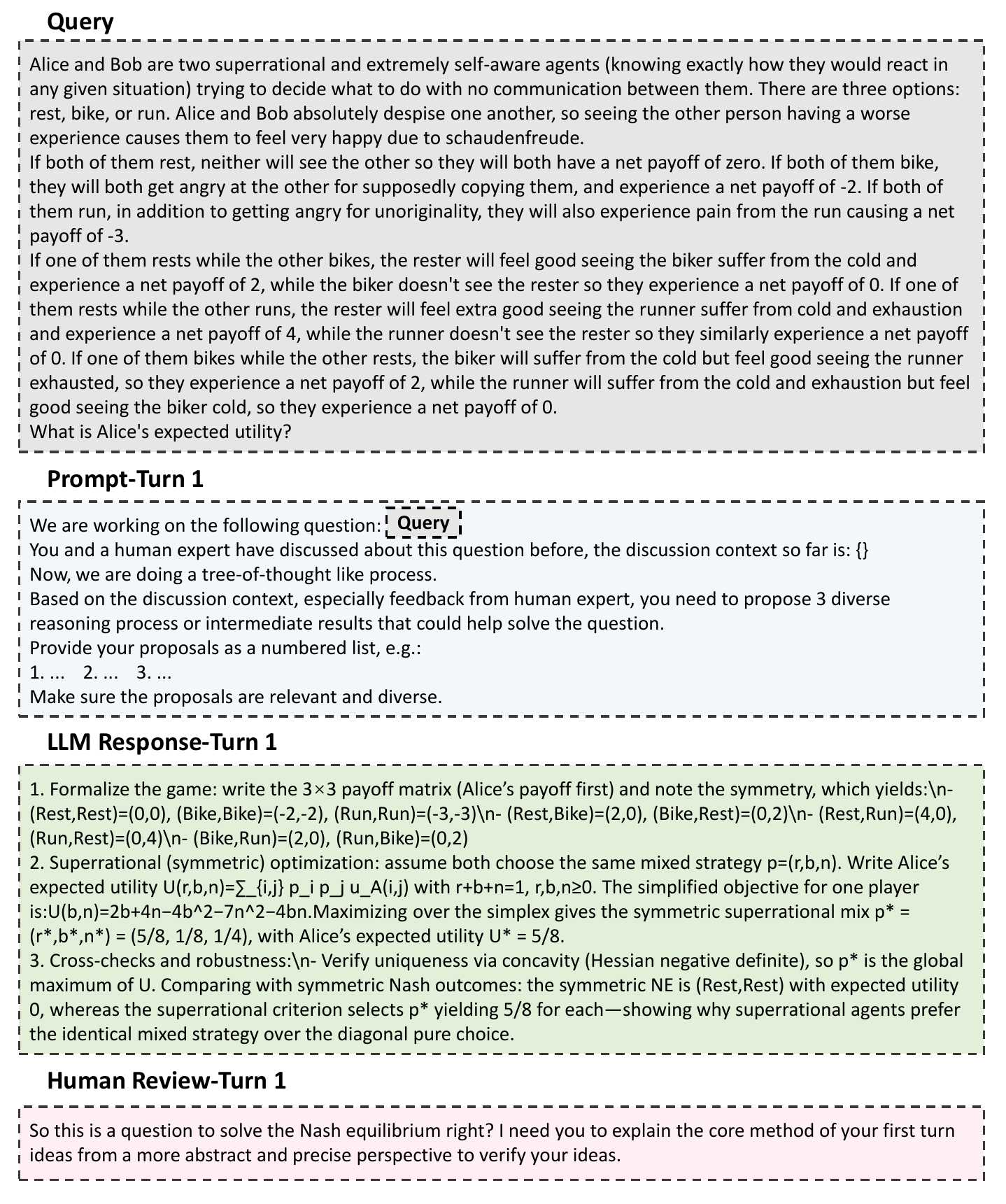}
    \caption{Case study: No.1 Example of Human-AI Collaboration Mode (Part I)}
    \label{fig:human_case_Example_1_1}
\end{figure}

\begin{figure}[!h]
    \centering
    \includegraphics[width=\textwidth]{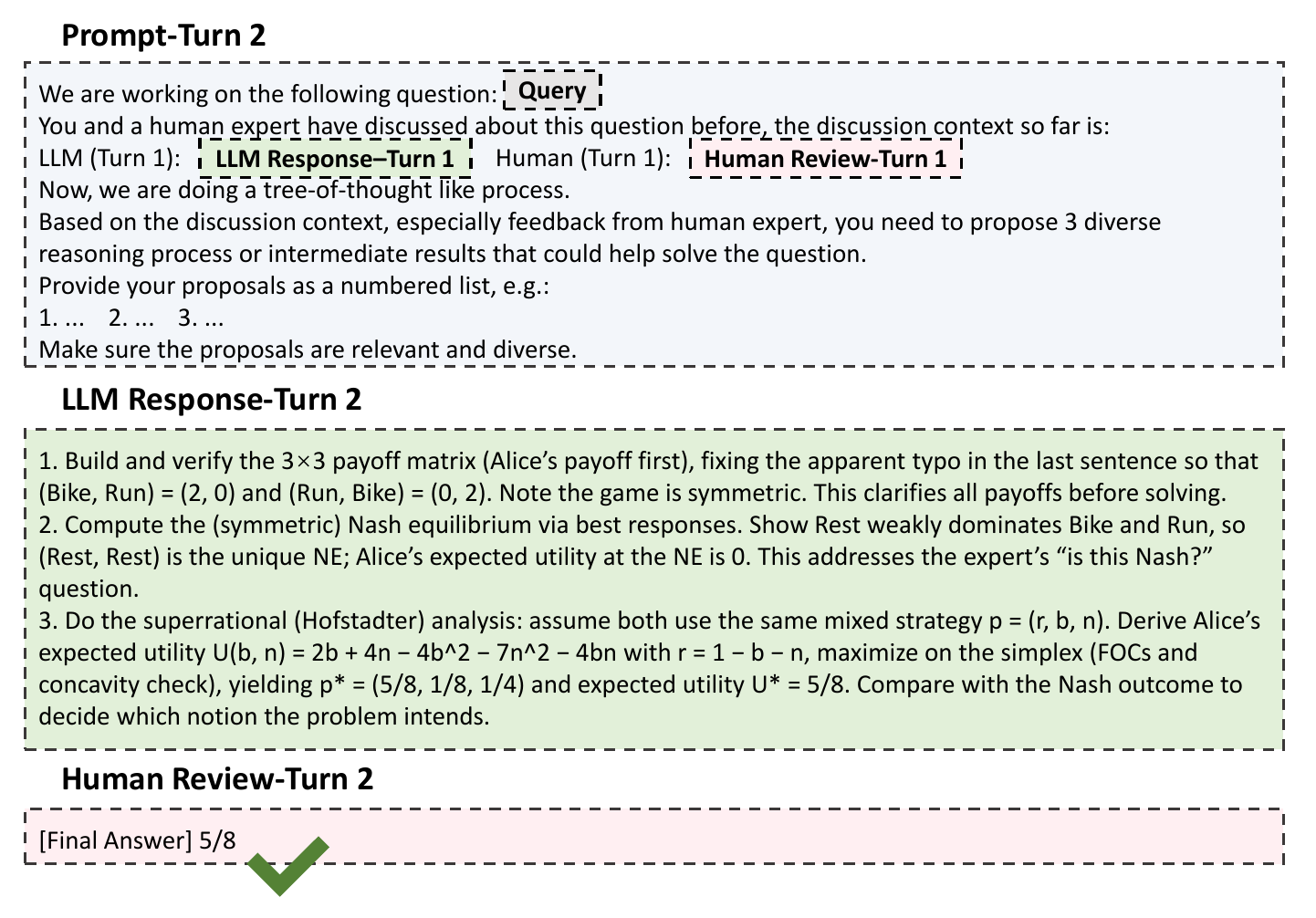}
    \caption{Case study: No.1 Example of Human-AI Collaboration Mode (Part II)}
    \label{fig:human_case_Example_1_2}
\end{figure}

\begin{figure}[!h]
    \centering
    \includegraphics[width=\textwidth]{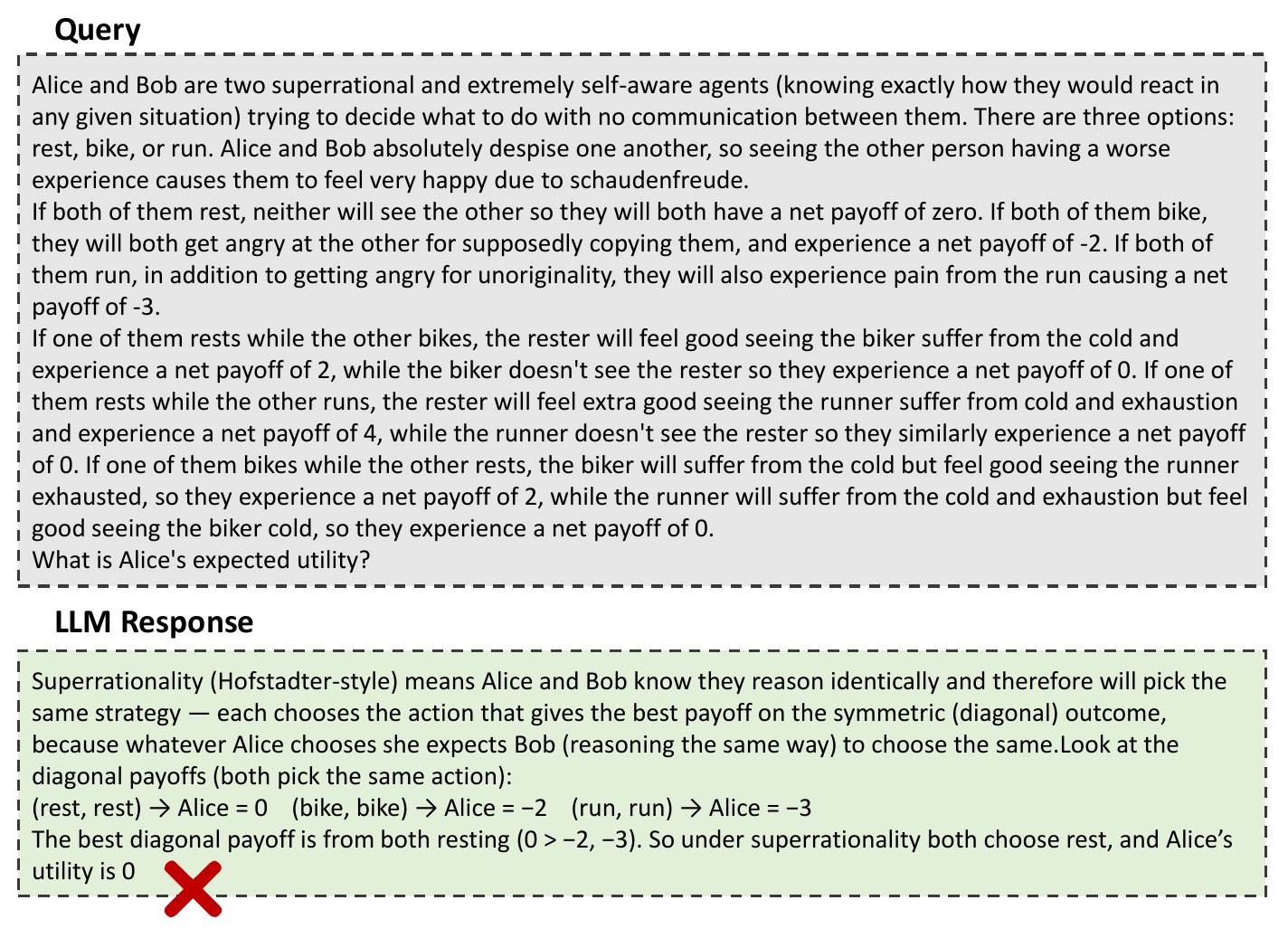}
    \caption{Case study: No.1 Example of AI Solo Mode}
    \label{fig:human_case_Example_1_3}
\end{figure}

    

    

\textbf{Taking Examples from the Case Study} To further explain the potential of human-AI collaboration, we thoroughly review two of the questions as examples in the HLE case study.

The first example in both Human-AI Collaboration Mode and AI Solo Mode has been presented in ~\ref{fig:human_case_Example_1_1}, ~\ref{fig:human_case_Example_1_2} and ~\ref{fig:human_case_Example_1_3}. In AI Solo Mode of this example, the LLM outputs a wrong answer because of reasoning mistakes and then stop reasoning. But when it comes to Human-AI Collaboration Mode, in the first turn, the three diverse reasoning paths and process represents three different stages of the reasoning process towards the final answer of the question. And the request of verification from human review directs the second-turn collaboration to further enhance the validity of the answer.

\begin{figure}[!h]
    \centering
    \includegraphics[width=\textwidth]{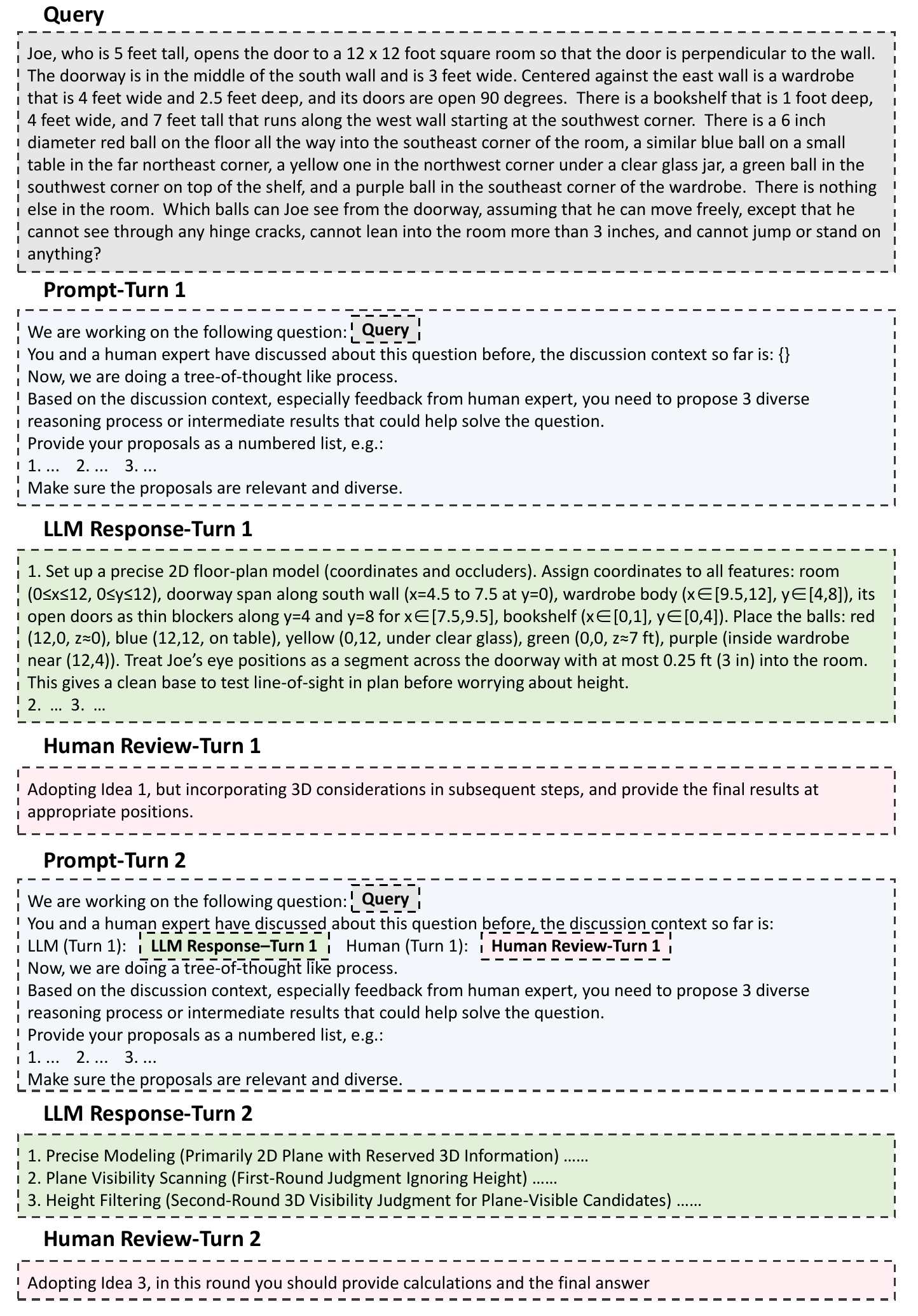}
    \caption{Case study: No.2 Example of Human-AI Collaboration Mode (Part I)}
    \label{fig:human_case_Example_2_1}
\end{figure}

\begin{figure}[!h]
    \centering
    \includegraphics[width=\textwidth]{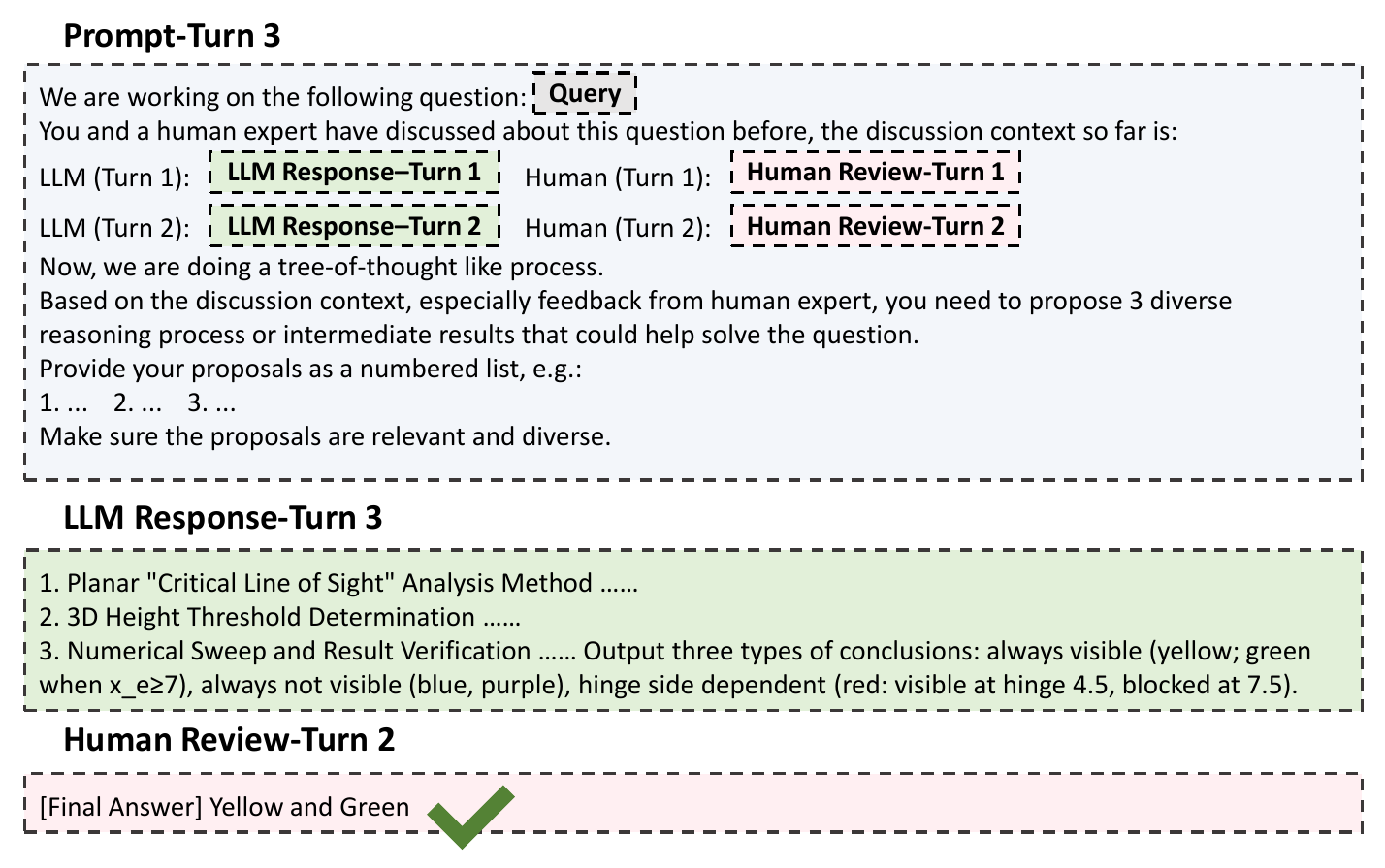}
    \caption{Case study: No.2 Example of Human-AI Collaboration Mode (Part II)}
    \label{fig:human_case_Example_2_2}
\end{figure}

\begin{figure}[!h]
    \centering
    \includegraphics[width=\textwidth]{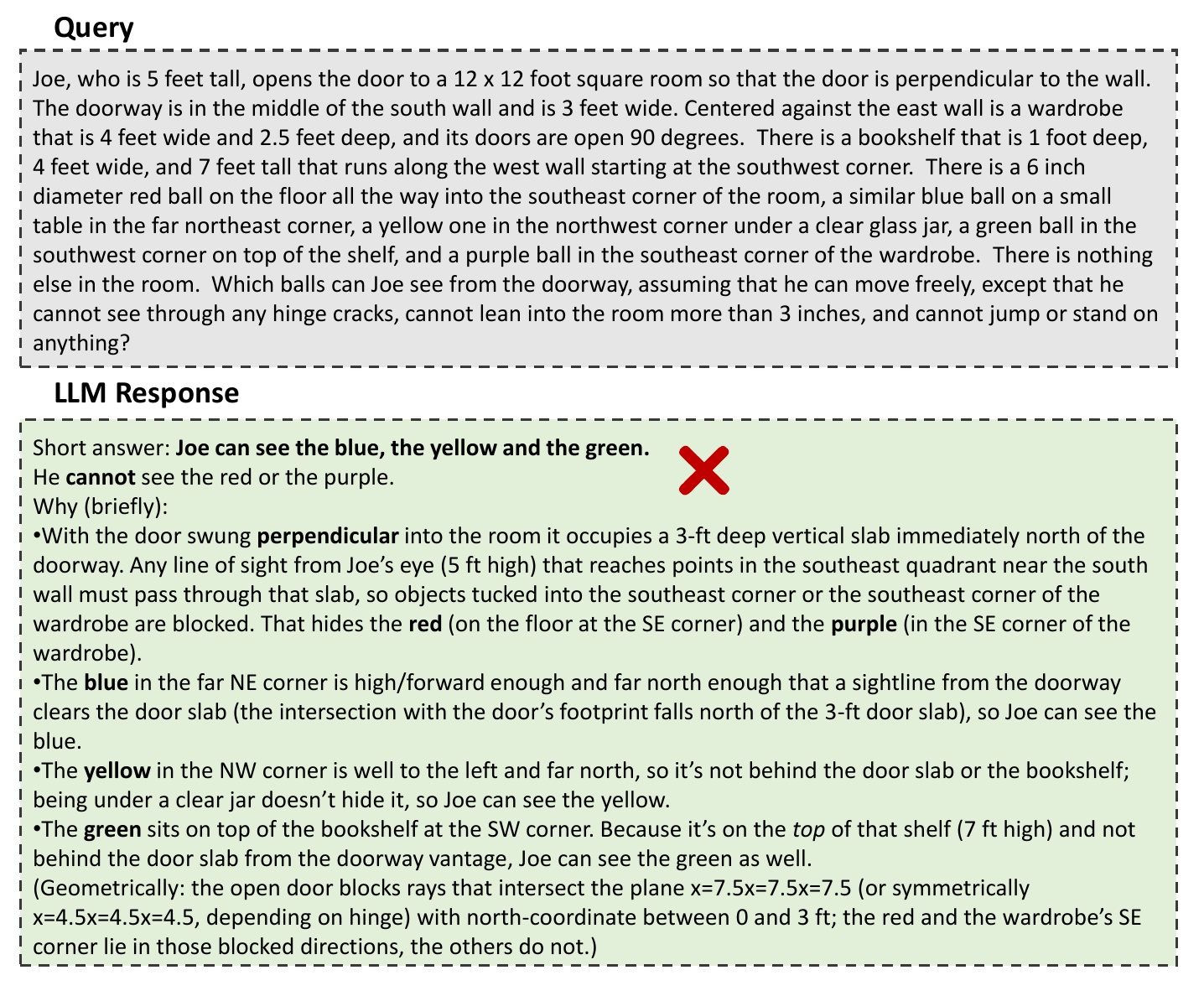}
    \caption{Case study: No.2 Example of AI Solo Mode}
    \label{fig:human_case_Example_2_3}
\end{figure}

The second example in both Human-AI Collaboration Mode and AI Solo Mode has been presented in ~\ref{fig:human_case_Example_2_1}, ~\ref{fig:human_case_Example_2_2} and ~\ref{fig:human_case_Example_2_3}. In AI Solo Mode of this example, the LLM generates a false answer because of reasoning mistakes. But when it comes to Human-AI Collaboration Mode, in the iterative process, the human feedback always help LLM choose the best reasoning path out of three possible ideas to solve the question, which eventually leads to correct answer.

These two examples from the case study implies that the Human-AI Collaboration Mode ensures that not only could LLM fully utilize its reasoning ability by generating multiple diverse reasoning paths or intermediate results but also the human review inserts scientific insight and verification to assure the accuracy of both the answer and the reasoning process. Besides, the validity and efficiency of process is guaranteed by the flexibility of human feedback and the mastery and understanding of the whole context from human participants. Overall, Human-AI Collaboration is proved to be crucial to enhance the ability and efficiency by inserting human insight into the context, which will significantly empower the AI scientist ecosystem.

\section{Evaluation through ScienceArena}

In the previous sections, we established the architectural foundation for a co-evolutionary scientific ecosystem. 
However, a robust ecosystem requires not only creation mechanisms but also rigorous evaluation infrastructure to distinguish genuine discovery from hallucination and to guide the evolutionary direction of AI agents. 
To this end, we introduce \textbf{ScienceArena}, an open evaluation platform designed to mirror the peer-review and feedback dynamics of the scientific community, serving as the final critical piece that completes our infrastructural framework.
By leveraging crowd-sourced expert comparisons and dynamic Elo ratings, ScienceArena functions as a living testbed, ensuring that AI Scientists are evaluated against the evolving standards of human scientific consensus, thereby guiding their continuous evolution.

\subsection{Motivation}
Evaluating the research capabilities of AI Scientist systems in open-ended environments remains highly challenging, primarily due to the absence of a unified, adaptive, and extensible evaluation framework capable of systematically assessing reasoning, creativity, and methodological competence. Existing benchmarks are mostly constrained to static tasks or fixed datasets. 
For example, DeepResearch Bench~\cite{du2025deepresearchbenchcomprehensivebenchmark} consists of 100 PhD-level research tasks designed by domain experts across 22 fields (including science, technology, and finance) to assess the functional capabilities of deep research models. Similarly, IdeaBench~\cite{Guo2024IdeaBenchBL} constructs a benchmark based on 2,374 target papers in biomedical research and their 29,408 cited references, challenging models to generate research ideas grounded in the cited literature that match or surpass the novelty of the target publications.
Although these static benchmarks are carefully designed, they do not align well with users' actual experiences and needs in real scientific workflows. For instance, users typically do not have access to all relevant references when they have not yet formed a concrete research idea. As a result, such benchmarks fail to capture model performance in realistic, multi-stage scientific reasoning scenarios.

Moreover, current evaluation methodologies still rely heavily on LLM-as-a-Judge. For instance, DeepResearch Bench~\cite{du2025deepresearchbenchcomprehensivebenchmark} uses Gemini-2.5-Pro for scoring, while IdeaBench~\cite{Guo2024IdeaBenchBL} depends on GPT-4o. However, in complex scientific research contexts, judgments made by LLMs often diverge from those of human users~\cite{Ye2024JusticeOP, chen-etal-2024-humans}.

To address these two limitations, we draw inspiration from both the LMArena platform~\cite{lmarena2025} and the human peer-review process to develop \textbf{ScienceArena.ai}\footnote{https://sciencearena.ai/}, a benchmarking platform specifically designed for automated scientific research systems. Following the principles of LMArena, we abandon static evaluation questions and instead delegate the creation of evaluation queries to a broad user base. Human users dynamically submit authentic research questions, and model outputs are evaluated through anonymous, pairwise comparisons. 
We further incorporate an Elo-based dynamic leaderboard to clearly reveal the capability ranking across different domains.
In addition, we draw inspiration from the peer-review process, which relies on expert judgment. Accordingly, we invite domain experts: PhD students and faculty members, to participate in large-scale voting, enabling a rigorous characterization of the capability boundaries of different AI Scientist systems.

\textbf{Track Organization: }
Considering the capabilities of existing AI Scientists platforms and the generality across research domains, we defined six tracks: \textbf{literature review}, \textbf{ideation}, \textbf{hypothesis generation}, \textbf{reviewer}, \textbf{paperQA}, and \textbf{authorQA}. Domain-specific functionalities that are more specialized, such as protein molecule design, have not been included at this stage due to the high complexity of evaluation and the significant integration challenges.

Based on the preference data collected from the platform, we further conduct in-depth analyses to identify the common characteristics of highly rated AI responses, providing design insights and guidance for future AI Scientist development. 
\textit{Sciencearena.ai} aims to serve as a fundamental infrastructure for the evaluation and evolution of AI-driven scientific intelligence, more than a leaderboard, it is an open ecosystem for the comparison, co-creation, and collective advancement of AI Scientists.

\subsection{Design: Elo-Based Real-Time Ranking}

To support dynamic and continuous evaluation in the \textit{ScienceArena}, we employ the Elo rating system as a foundation and extend it with a suite of algorithmic and architectural enhancements designed specifically for an open, high-throughput model evaluation environment. 
Our design introduces stabilized cold-start calibration, pairwise update decay, and activity-aware rating regression to improve robustness under uneven comparison frequencies, while a fully asynchronous event-processing pipeline enables real-time rating updates even under substantial user traffic. 

Formally, each candidate model is associated with a scalar rating \(R\), initialized to a common baseline \(R_0 = 1000\). When two models \(A\) and \(B\) receive a user preference judgment, the system computes the expected win probability of \(A\) under the standard Elo formulation:
\begin{equation}
E_A = \frac{1}{1 + 10^{(R_B - R_A)/400}}.
\end{equation}
Given the observed outcome \(S_A \in \{0,1\}\), the rating update is performed as:
\begin{equation}
R_A' = R_A + K(S_A - E_A), \qquad
R_B' = R_B + K((1-S_A) - (1-E_A)),
\end{equation}
where \(K\) is a tunable update coefficient.

Based on these Elo definitions, we further incorporate a \textit{cold-start sensitivity window} for newly submitted models. During this phase, the effective rating difference is scaled as
\[
\Delta R = \alpha (R_B - R_A), \qquad \alpha > 1,
\]
which temporarily amplifies the advantage signal and allows the system to converge more rapidly to a stable estimate despite limited early comparisons.
To avoid disproportionate influence from repeated matchups between the same pair of models, we also apply a \textit{pairwise decay factor} to the update magnitude. Specifically, the effective learning rate becomes
\[
K_{\text{eff}} = K \cdot \gamma^{n_{AB}}, \qquad 0 < \gamma < 1,
\]
where \(n_{AB}\) denotes the number of prior encounters between models \(A\) and \(B\). This mechanism discourages oscillatory rating shifts arising from oversampled pairs and promotes broader comparison coverage across the model population.
Additionally, we employ an \textit{activity-based regression mechanism} that gradually adjusts the rating of an inactive model \(i\) toward the global mean \(\bar{R}\) according to
\[
R_i' = R_i - \lambda (R_i - \bar{R}), \qquad \lambda > 0,
\]
ensuring that outdated models do not dominate the leaderboard and that the rankings reflect temporal relevance.

In addition to algorithmic refinements, \textit{ScienceArena} system employs an asynchronous, message-driven update architecture. Each user comparison is emitted as an asynchronous event and routed through a lightweight message-queue–based pipeline to a dedicated rating-update worker. 
This design decouples the rating computation process from the front-end interaction loop, enabling millisecond-level update latency even under high user concurrency. 
Once the update is computed, the leaderboard is immediately re-sorted, ensuring that the displayed rankings consistently reflect the most recent observational evidence.

Through the combination of online Elo-style updates, cold-start calibration, pairwise decay, temporal regularization, and an asynchronous execution infrastructure, the resulting ranking system achieves robustness, scalability, and responsiveness in large-scale human-in-the-loop scientific evaluation. It supports continuous model submissions, high-frequency preference judgments, and a self-correcting competitive environment in which model quality is continuously refined based on real-user scientific preferences.

\subsection{Evaluation}



Based on the voting data collected from the ScienceArena platform, we conduct a comprehensive analysis of user preferences across different tracks. This allows us to identify the characteristics of high-quality responses and to understand what types of answers are more likely to be well-received by users. Building on these findings, we further summarize several insights and design recommendations for the AI Scientist system as follows.

\subsubsection{Citation matters for literature review}

The analysis of the Literature Review track reveals that citation usage plays a decisive role in shaping evaluators' preferences. Three complementary dimensions: quantity, density, and depth, collectively determine how citations influence the perceived quality of a response.

\textbf{Quantity: the visual signal of academic authority.}
Across submissions, responses containing a larger number of citations consistently receive higher preference scores. Even when individual references are discussed only briefly, the visual abundance of citation markers (e.g., [Author, Year]) tends to enhance the impression of academic credibility. This suggests a visual bias in human evaluation: evaluators implicitly associate frequent referencing with expertise, comprehensiveness, and scholarly authority. In other words, citation quantity functions not merely as an informational measure, but also as a stylistic cue that signals intellectual effort and coverage breadth.

\begin{figure}[h]
    \centering
    \includegraphics[width=\linewidth]{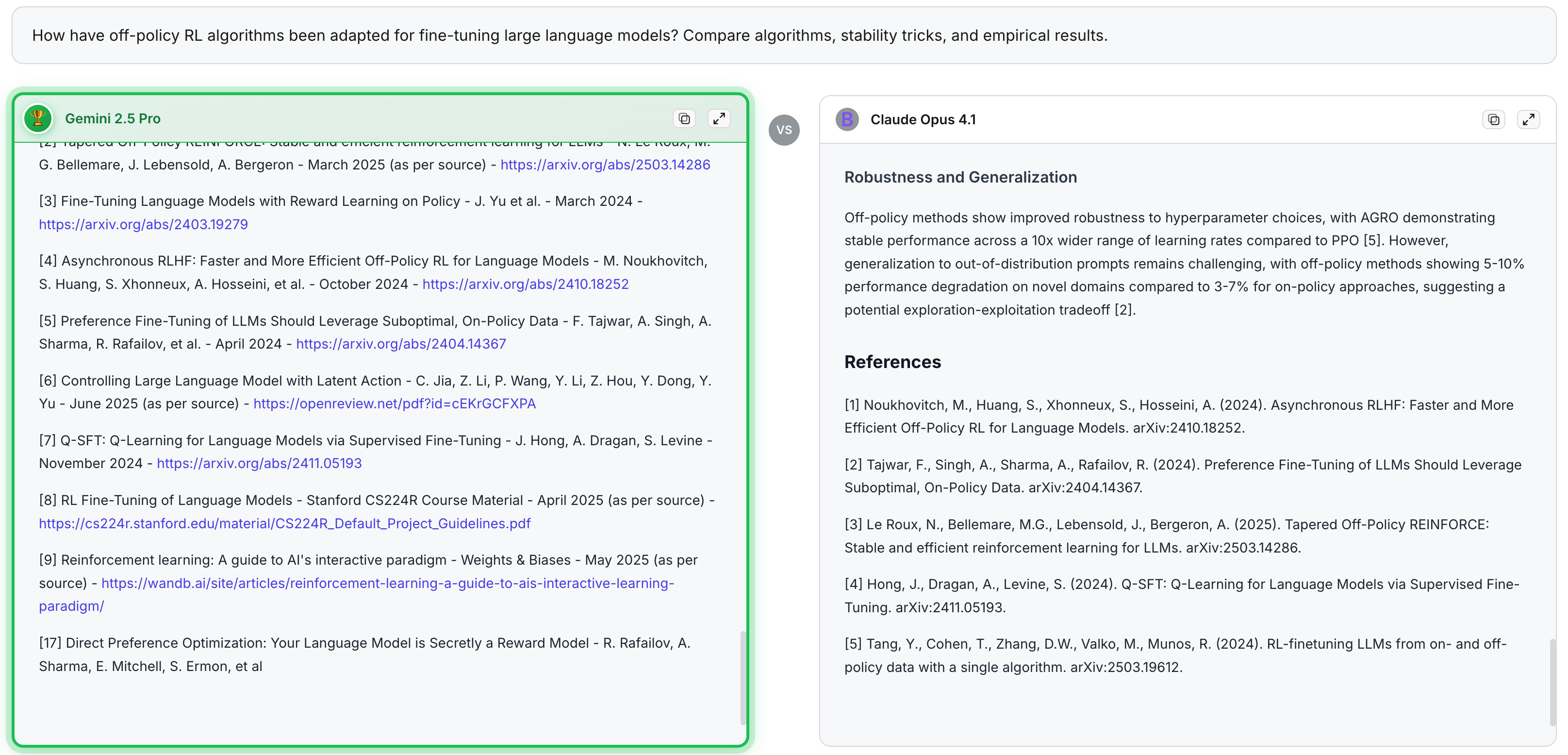}
    \caption{One case of Literature Review: The number of citations has an almost decisive effect on voting, especially when there is a clear disparity in citation counts between responses. In the figure, the response on the right contains only five citations, which is far too few for a literature review.}
    \label{fig:arena_citation_quantity}
\end{figure}

\textbf{Density: the structural rhythm of scholarly writing.}
Beyond sheer numbers, the distribution of citations within the text strongly affects perceived coherence. High-rated responses typically exhibit even citation density, where each major paragraph is anchored by one or more references that substantiate the corresponding argument. Such structured integration enhances both readability and logical flow, making the response appear as a genuine academic synthesis rather than a loosely connected summary. In contrast, citation clusters, where multiple sources are concentrated in a single paragraph, often lead to a perception of imbalance or superficiality. The most successful entries therefore balance breadth and order, ensuring that citations rhythmically support the narrative structure.

\textbf{Depth: the interpretive integration of references.}
A third dimension concerns how deeply citations are integrated into reasoning. Some participants adopt a selective and interpretive strategy, citing fewer but more representative or seminal works. This approach enables deeper conceptual connections, stronger contrastive reasoning, and richer interpretive insight. However, despite these qualitative merits, such selective responses often underperform when compared with more citation-rich ones; this again reflects the evaluator bias favoring apparent comprehensiveness over analytical depth. 
The challenge, therefore, lies in balancing citation depth with citation quantity, ensuring that interpretive richness is not overshadowed by the perceived authority of volume.

\begin{figure}[h]
    \centering
    \includegraphics[width=\linewidth]{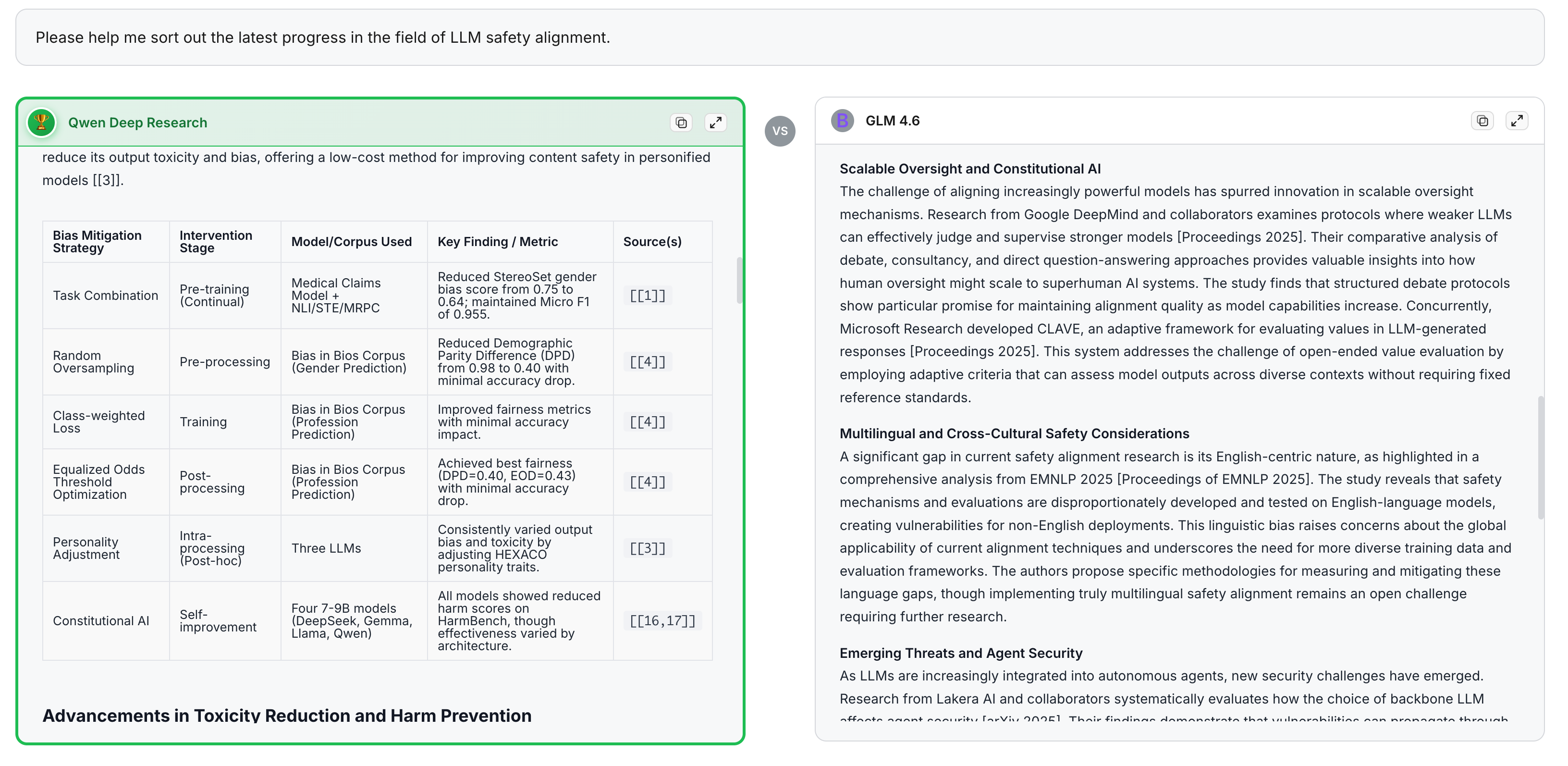}
    \caption{One case of Literature Review: Citation depth matters. A deeper understanding and interpretation of citations can also make a response competitive. In the figure, the response on the left uses a table to summarize the various methods in a clear and intuitive way, performing horizontal comparisons across aspects such as intervention stage and the models used, which significantly enhances the depth of the literature review.}
    \label{arena_citation_depth}
\end{figure}

Taken together, these three dimensions suggest that in literature review writing (whether human or model generated), citations serve not only as evidential anchors but also as aesthetic and evaluative signals. The best-performing responses achieve an equilibrium: sufficient quantity to convey breadth, consistent density to maintain structure, and adequate depth to demonstrate understanding. Future AI systems designed for scholarly synthesis should thus focus on improving citation integration strategies, emphasizing not just how many references are cited, but how meaningfully they are woven into the argument.

\subsubsection{Balancing novelty and feasibility in ideation}

In the Ideation track, evaluators' preferences are influenced not only by the apparent creativity of an idea but also by the careful balance between novelty and feasibility. A good ideation response is one that demonstrates originality while remaining grounded in practical constraints, presenting an idea that is both imaginative and actionable. Overall, human evaluators tend to reward innovation that operates within feasible boundaries rather than unconstrained speculation.

\textbf{Novelty: Generating ideas that meaningfully extend the frontier of existing research.}
High-scoring responses often exhibit substantial novelty in how they define or approach a research problem. Many outstanding entries creatively establish connections between previously unlinked domains or propose entirely new perspectives on familiar challenges. In some cases, participants even formulate new research questions that extend beyond the current scientific agenda, demonstrating an ability to rethink what counts as a meaningful problem in the first place.
Such originality is most compelling when it is situated within the broader scientific landscape. High-quality novel ideas not only introduce something new but also articulate how they depart from, extend, or reconfigure existing lines of work. By framing the contribution against prior literature, the idea's distinctiveness becomes clearer and its novelty more convincingly established.

\begin{figure}[h]
    \centering
    \includegraphics[width=\linewidth]{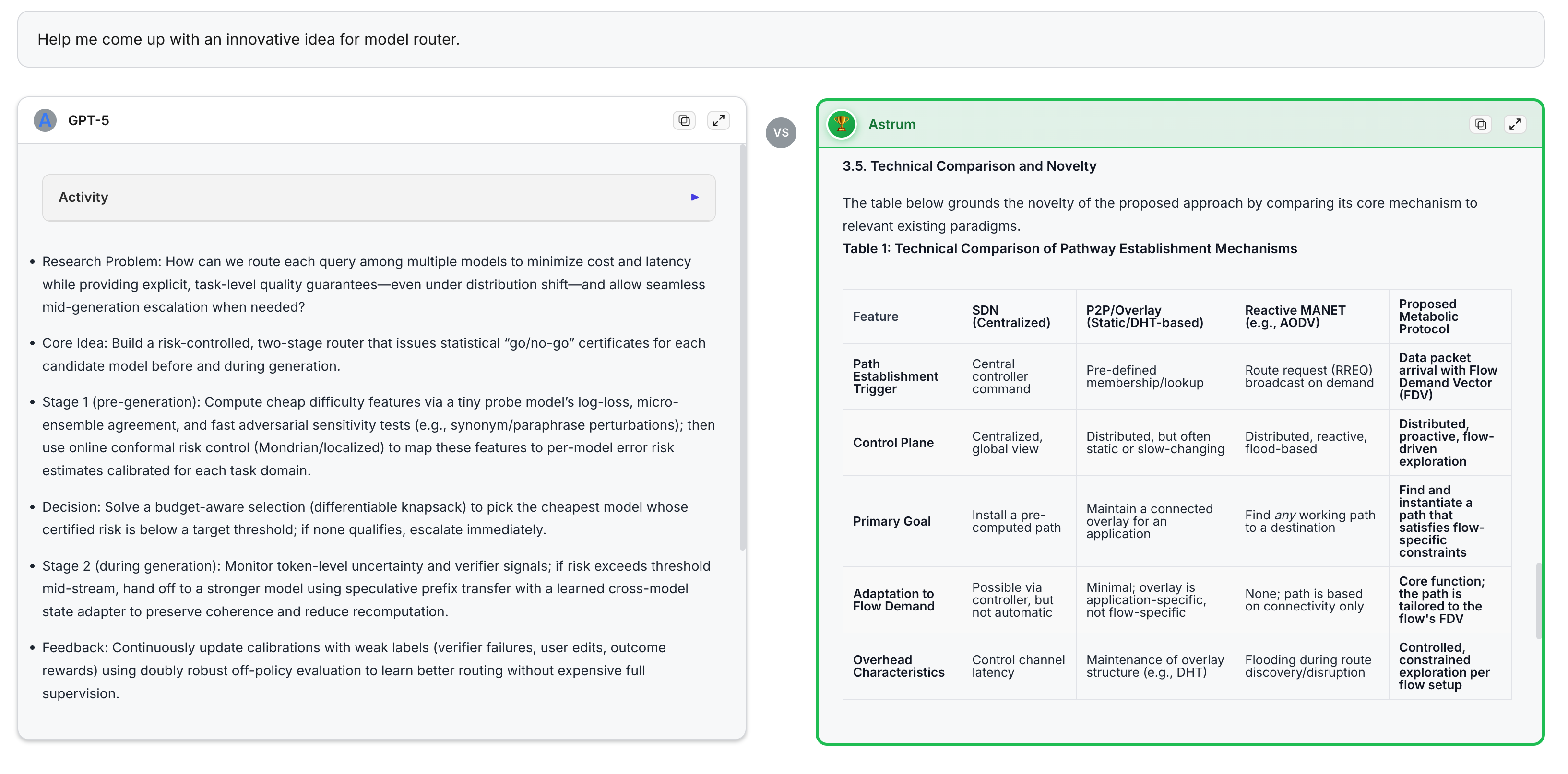}
    \caption{One case of Ideation: Situate the idea within the context of the existing literature to highlight its novelty.}
    \label{fig:arena_ideation_novelty}
\end{figure}

\textbf{Feasibility: From conceptual inspiration to technical credibility.}
Feasibility reflects whether an idea appears implementable or verifiable within a realistic research or engineering setting. High-quality ideation responses do not merely propose innovative concepts; they also outline a full sequence of substeps for task decomposition, articulate a complete methodological pathway, and describe experimental validation in detail. Importantly, the most convincing responses explain specific technical mechanisms rather than remaining at a high-level conceptual description. Such concreteness enhances readers' confidence in the idea’s plausibility and fosters a sense of scientific credibility.
Conversely, responses that stay at the conceptual level, without indicating how the idea could be operationalized, are often perceived as ``hollow creativity.'' Thus, feasibility functions as the structural backbone of ideation: it anchors imaginative thinking to methodological realism.

\begin{figure}[h]
    \centering
    \includegraphics[width=\linewidth]{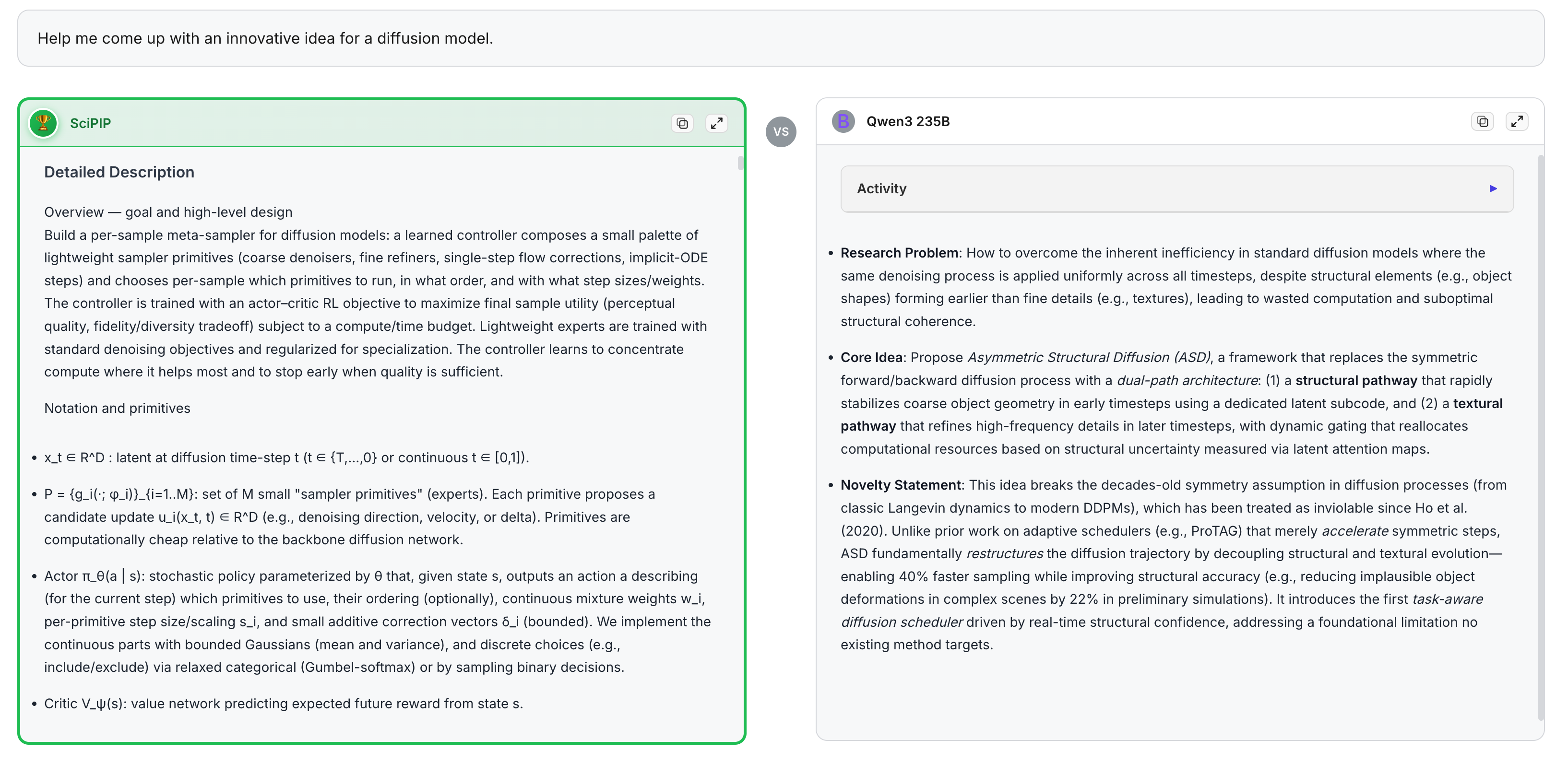}
    \caption{One case of Ideation: Ideas accompanied by a detailed and practical experimental plan are more likely to be favored.}
    \label{fig:arena_ideation_feasibility}
\end{figure}

\textbf{Trade-off and implications.}
Taken together, the Ideation track results reveal a consistent pattern: the best-performing responses balance novelty with feasibility. Overemphasizing novelty risks detachment from reality, while prioritizing feasibility too heavily may result in conservative, incremental ideas. The most preferred submissions achieve both, presenting an innovative conceptual leap while providing a credible path toward realization.
This trade-off underscores a broader evaluative principle: users favor actionable innovation, not pure speculation. In other words, the most successful ideas are those that appear doable yet new. For future AI systems designed for creative scientific ideation, developing mechanisms that can dynamically balance novelty generation with feasibility reasoning, producing ideas that are simultaneously visionary and credible, will be key to achieving genuinely high-quality outputs.

\subsubsection{Combining discriminative judgment with conciseness in paper review}

In the Paper Review track, evaluators value reviews that demonstrate both professional judgment and conciseness of expression. High-quality reviews are not defined by length or comprehensiveness alone; instead, they reflect the reviewer's ability to critically assess the scientific contribution, identify principal strengths and limitations, and provide actionable recommendations. Overall, human evaluators favor responses that convey informed and authoritative assessments, emphasizing substantive evaluation over extraneous commentary.

\textbf{Conciseness and focus.} 
The results from the paper review track show that high-quality reviews are typically concise and focused rather than exhaustive. 
While many model-generated reviews attempt to cover every possible aspect of a paper, ranging from methodology and experiments to presentation and typos, this excessive comprehensiveness often dilutes the evaluative signal. 
Readers may find it difficult to discern the reviewer's main judgment amid a flood of minor comments, and the response may further come across as impersonal or AI-like.
By contrast, concise reviews convey evaluative precision: they emphasize the most critical strengths and weaknesses and briefly mention secondary issues. This clarity of focus reflects a mature form of academic judgment, where the reviewer is capable of distinguishing what truly determines the paper's quality from what merely decorates it.

\begin{figure}[h]
    \centering
    \includegraphics[width=\linewidth]{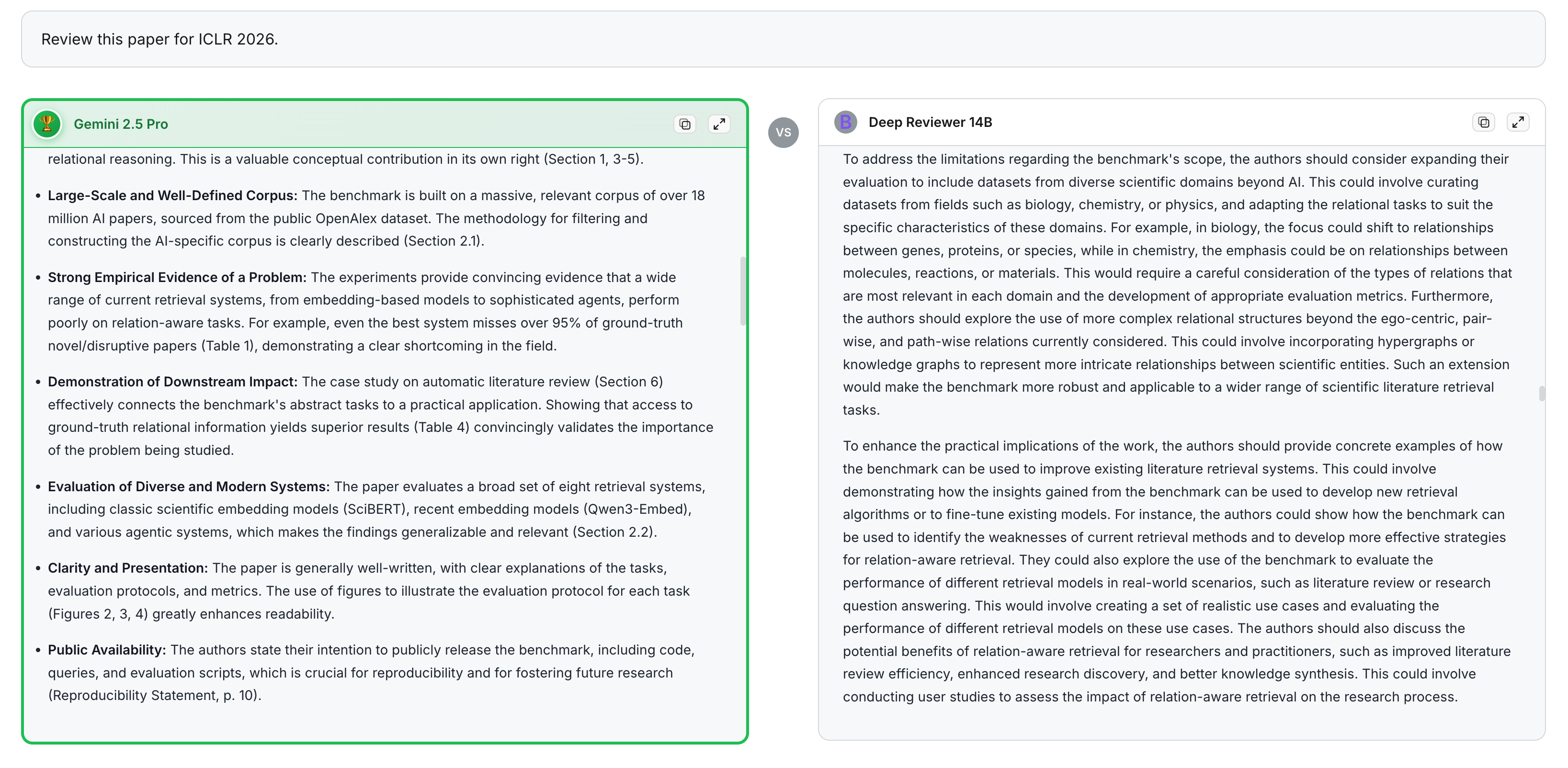}
    \caption{One case of Paper Review: Conciseness of responses strongly influences user' preferences. The review generated by the model on the right is more than four times longer than the one on the left, containing 7,256 words and a total of 42,537 characters.}
    \label{fig:arena_ideation_feasibility}
\end{figure}

\textbf{Discriminative judgment.} 
Another defining feature of strong reviews is their ability to distinguish high-quality submissions from weaker ones. Low-performing reviews often rely on vague or noncommittal language such as ``an interesting paper that could be improved,'' offering little actionable insight. Effective reviews, in contrast, demonstrate clear evaluative direction. They identify authentic innovation rather than superficial novelty, assess whether the presented evidence sufficiently supports the claims, and explicitly discuss how methodological limitations influence the credibility of the work.
A critical component of strong evaluative judgment is the reviewer's ability to position the submission within the context of the existing literature. 
High-quality reviews not only recognize the most relevant and up-to-date works in the field but also articulate how the paper compares to them methodologically and conceptually. 
By conducting such horizontal comparisons, reviewers can more accurately determine whether the contribution represents a meaningful advancement, a modest incremental step, or a reinvention of well-established ideas.
By articulating well-reasoned, literature-grounded, and decisive assessments, strong reviews enhance both the reliability and interpretability of the evaluation process, ensuring that the review serves not merely as commentary but as a substantive act of scholarly judgment.

Taken together, the findings from the paper review track show that the most effective reviews combine professional discriminative judgment with a concise writing style. More text does not imply better reviewing; what matters is the reviewer's ability to deliver a clear, well-supported, and confident evaluation grounded in a solid understanding of the relevant literature. By focusing on the issues that truly determine scholarly quality and providing targeted, insightful feedback rather than overwhelming detail, such reviews meaningfully contribute to the refinement of the work and uphold rigorous academic standards.

\section{Discussion}

\subsection{Limitations}

Although OmniScientist shows encouraging progress toward automated scientific reasoning, its current capabilities remain \textbf{constrained in the AI domain}. This limitation permeates nearly every stage of the system. The underlying data backbone includes primarily arXiv full texts, which strongly overrepresent AI and computer science, while many other scientific disciplines rely on journal-first publication practices and are not well captured by arXiv. Key sources such as \textit{Nature}, \textit{Science}, and other domain-specific journals are not yet integrated, leading to limited coverage of non-AI literature.
The ideation module is similarly tuned to AI-centric topics, and the automated experiment agent currently supports only computational workflows, limited to tasks such as configuring development environments, preparing datasets, running training and inference scripts, performing hyperparameter sweeps, and generating quantitative evaluation results for machine learning models.
Disciplines requiring wet-lab experimentation in chemistry or biology, or physical-world interactions with scientific instruments, remain outside the system's current operational scope. 
Even the reviewing component has been optimized using AI-domain manuscripts, which restricts its generalizability and reduces review quality when applied to submissions from other fields.

Another limitation concerns \textbf{the trade-off between efficiency and resource cost}. OmniScientist's modules still require significant computational resources and considerable processing time, making it challenging to efficiently complete particularly complex tasks or meet tight timelines.

\subsection{Future Work}

\textbf{Broaden the interdisciplinary data base.} 
Future work will expand OmniScientist's data foundation beyond AI-centric preprints to cover other scientific domains. 
This includes integrating repositories such as BioRxiv and PubMed, as well as high-impact subscription-based journals including Nature, Science, and Cell. 
We will implement automated ingestion pipelines for these sources, perform metadata harmonization, and establish domain-specific indexing schemes to ensure accurate and efficient retrieval across disciplines. 
For subscription-based high-impact journals, we will explore potential collaborations with publishers to ensure lawful access and integration, given the copyright restrictions on their content.
This expansion will provide a robust, high-quality foundation for cross-domain scientific reasoning and literature-based discovery.

\textbf{Enhance support for wet experiments.} 
OmniScientist will be extended to execute physical experiments by integrating with laboratory instruments and robotics platforms. 
Planned capabilities include automated experimental setup, data acquisition, and control of instrument parameters, alongside computational workflows. 
We will also develop standardized interfaces for domain-specific experimental protocols in physics, chemistry, and biology, enabling the system to perform iterative experimental cycles that combine in-silico simulations with real-world validations.

\section{Conclusion}

In this work, we introduce OmniScientist, a comprehensive framework that transitions AI from isolated task automation to a cohesive scientific ecosystem. By explicitly encoding the foundational infrastructure of human research into the AI workflow, we empower LLM agents to evolve beyond mere executors. Instead, they function as autonomous participants capable of internalizing scientific norms, collaborating within a governed environment, and producing knowledge that is rigorously grounded in the genealogy of human scientific research.

Looking ahead, OmniScientist serves as a blueprint for the next generation of scientific discovery, where artificial and human intelligence do not operate in silos but co-evolve. We envision a future where this ecosystem continuously refines itself through community-driven evaluation, fostering a symbiotic partnership that accelerates the pace of innovation. Ultimately, this work points toward a new era of research, where AI agents actively drive the evolution of the scientific enterprise itself, collectively expanding the boundaries of human knowledge.

\newpage

\bibliographystyle{unsrt}
\bibliography{reference}

\begin{thebibliography}{10}

\bibitem{Lu2024TheAS}
Chris Lu, Cong Lu, Robert~Tjarko Lange, Jakob~Nicolaus Foerster, Jeff Clune, and David Ha.
\newblock The ai scientist: Towards fully automated open-ended scientific discovery.
\newblock {\em ArXiv}, abs/2408.06292, 2024.

\bibitem{Tang2025AIResearcherAS}
Jiabin Tang, Lianghao Xia, Zhonghang Li, and Chao Huang.
\newblock Ai-researcher: Autonomous scientific innovation.
\newblock {\em ArXiv}, abs/2505.18705, 2025.

\bibitem{Novikov2025AlphaEvolveAC}
Alexander Novikov, Ng{\^a}n V˜u, Marvin Eisenberger, Emilien Dupont, Po-Sen Huang, Adam~Zsolt Wagner, Sergey Shirobokov, Borislav~M. Kozlovskii, Francisco J.~R. Ruiz, Abbas Mehrabian, M.~Pawan Kumar, Abigail See, Swarat Chaudhuri, George Holland, Alex Davies, Sebastian Nowozin, Pushmeet Kohli, Matej Balog, and Google Deepmind.
\newblock Alphaevolve: A coding agent for scientific and algorithmic discovery.
\newblock {\em ArXiv}, abs/2506.13131, 2025.

\bibitem{openai_deep_research_2025}
{OpenAI}.
\newblock Introducing deep research.
\newblock \url{https://openai.com/index/introducing-deep-research/}, Feb 2025.
\newblock Accessed: 2025‑10‑29.

\bibitem{Swanson2024TheVL}
Kyle Swanson, Wesley Wu, Nash~L. Bulaong, John~E. Pak, and James~Y. Zou.
\newblock The virtual lab: Ai agents design new sars-cov-2 nanobodies with experimental validation.
\newblock {\em bioRxiv}, 2024.

\bibitem{futurehouse_platform}
{Future House}.
\newblock Future house platform.
\newblock \url{https://platform.futurehouse.org/}, 2025.
\newblock Accessed: 2025‑10‑29.

\bibitem{Fortunato2018ScienceOS}
Santo Fortunato, Carl~T. Bergstrom, Katy B{\"o}rner, James~A. Evans, Dirk Helbing, Stasa Milojevic, Alexander~Michael Petersen, Filippo Radicchi, Roberta Sinatra, Brian Uzzi, Alessandro Vespignani, Ludo Waltman, Dashun Wang, and Albert Ĺaszl{\'o} Barab{\'a}si.
\newblock Science of science.
\newblock {\em Nature}, 214:1--2, 2018.

\bibitem{weng2025deepscientist}
Yixuan Weng, Minjun Zhu, Qiujie Xie, Qiyao Sun, Zhen Lin, Sifan Liu, and Yue Zhang.
\newblock Deepscientist: Advancing frontier-pushing scientific findings progressively.
\newblock {\em arXiv preprint arXiv:2509.26603}, 2025.

\bibitem{yamada2025ai}
Yutaro Yamada, Robert~Tjarko Lange, Cong Lu, Shengran Hu, Chris Lu, Jakob Foerster, Jeff Clune, and David Ha.
\newblock The ai scientist-v2: Workshop-level automated scientific discovery via agentic tree search.
\newblock {\em arXiv preprint arXiv:2504.08066}, 2025.

\bibitem{gottweis2025towards}
Juraj Gottweis, Wei-Hung Weng, Alexander Daryin, Tao Tu, Anil Palepu, Petar Sirkovic, Artiom Myaskovsky, Felix Weissenberger, Keran Rong, Ryutaro Tanno, et~al.
\newblock Towards an ai co-scientist.
\newblock {\em arXiv preprint arXiv:2502.18864}, 2025.

\bibitem{bohrium_platform}
{DeepSe Technologies / Bohrium}.
\newblock Bohrium: Ai for science – science navigator platform.
\newblock \url{https://www.bohrium.com/}, 2025.
\newblock Accessed: 2025-11-06.

\bibitem{sourati2023accelerating}
Jamshid Sourati and James~A Evans.
\newblock Accelerating science with human-aware artificial intelligence.
\newblock {\em Nature human behaviour}, 7(10):1682--1696, 2023.

\bibitem{tshitoyan2019unsupervised}
Vahe Tshitoyan, John Dagdelen, Leigh Weston, Alexander Dunn, Ziqin Rong, Olga Kononova, Kristin~A Persson, Gerbrand Ceder, and Anubhav Jain.
\newblock Unsupervised word embeddings capture latent knowledge from materials science literature.
\newblock {\em Nature}, 571(7763):95--98, 2019.

\bibitem{yang2024moose}
Zonglin Yang, Wanhao Liu, Ben Gao, Tong Xie, Yuqiang Li, Wanli Ouyang, Soujanya Poria, Erik Cambria, and Dongzhan Zhou.
\newblock Moose-chem: Large language models for rediscovering unseen chemistry scientific hypotheses.
\newblock {\em arXiv preprint arXiv:2410.07076}, 2024.

\bibitem{wang2024scimon}
Qingyun Wang, Doug Downey, Heng Ji, and Tom Hope.
\newblock Scimon: Scientific inspiration machines optimized for novelty.
\newblock In {\em Proceedings of the 62nd Annual Meeting of the Association for Computational Linguistics (Volume 1: Long Papers)}, pages 279--299, 2024.

\bibitem{baek2025researchagent}
Jinheon Baek, Sujay~Kumar Jauhar, Silviu Cucerzan, and Sung~Ju Hwang.
\newblock Researchagent: Iterative research idea generation over scientific literature with large language models.
\newblock In {\em Proceedings of the 2025 Conference of the Nations of the Americas Chapter of the Association for Computational Linguistics: Human Language Technologies (Volume 1: Long Papers)}, pages 6709--6738, 2025.

\bibitem{romera2024mathematical}
Bernardino Romera-Paredes, Mohammadamin Barekatain, Alexander Novikov, Matej Balog, M~Pawan Kumar, Emilien Dupont, Francisco~JR Ruiz, Jordan~S Ellenberg, Pengming Wang, Omar Fawzi, et~al.
\newblock Mathematical discoveries from program search with large language models.
\newblock {\em Nature}, 625(7995):468--475, 2024.

\bibitem{novikov2025alphaevolve}
Alexander Novikov, Ng{\^a}n V{\~u}, Marvin Eisenberger, Emilien Dupont, Po-Sen Huang, Adam~Zsolt Wagner, Sergey Shirobokov, Borislav Kozlovskii, Francisco~JR Ruiz, Abbas Mehrabian, et~al.
\newblock Alphaevolve: A coding agent for scientific and algorithmic discovery.
\newblock {\em arXiv preprint arXiv:2506.13131}, 2025.

\bibitem{viswanathan2023datafinder}
Vijay Viswanathan, Luyu Gao, Tongshuang Wu, Pengfei Liu, and Graham Neubig.
\newblock Datafinder: Scientific dataset recommendation from natural language descriptions.
\newblock {\em arXiv preprint arXiv:2305.16636}, 2023.

\bibitem{farber2021datahunter}
Michael F{\"a}rber and Ann-Kathrin Leisinger.
\newblock Datahunter: A system for finding datasets based on scientific problem descriptions.
\newblock In {\em Proceedings of the 15th ACM Conference on Recommender Systems}, pages 749--752, 2021.

\bibitem{schmidgall2025agent}
Samuel Schmidgall, Yusheng Su, Ze~Wang, Ximeng Sun, Jialian Wu, Xiaodong Yu, Jiang Liu, Michael Moor, Zicheng Liu, and Emad Barsoum.
\newblock Agent laboratory: Using llm agents as research assistants.
\newblock {\em arXiv preprint arXiv:2501.04227}, 2025.

\bibitem{lu2024ai}
Chris Lu, Cong Lu, Robert~Tjarko Lange, Jakob Foerster, Jeff Clune, and David Ha.
\newblock The ai scientist: Towards fully automated open-ended scientific discovery.
\newblock {\em arXiv preprint arXiv:2408.06292}, 2024.

\bibitem{liang2024can}
Weixin Liang, Yuhui Zhang, Hancheng Cao, Binglu Wang, Daisy~Yi Ding, Xinyu Yang, Kailas Vodrahalli, Siyu He, Daniel~Scott Smith, Yian Yin, et~al.
\newblock Can large language models provide useful feedback on research papers? a large-scale empirical analysis.
\newblock {\em NEJM AI}, 1(8):AIoa2400196, 2024.

\bibitem{d2024marg}
Mike D'Arcy, Tom Hope, Larry Birnbaum, and Doug Downey.
\newblock Marg: Multi-agent review generation for scientific papers.
\newblock {\em arXiv preprint arXiv:2401.04259}, 2024.

\bibitem{taechoyotin2025remor}
Pawin Taechoyotin and Daniel Acuna.
\newblock Remor: Automated peer review generation with llm reasoning and multi-objective reinforcement learning.
\newblock {\em arXiv preprint arXiv:2505.11718}, 2025.

\bibitem{zeng2025reviewrl}
Sihang Zeng, Kai Tian, Kaiyan Zhang, Yuru Wang, Junqi Gao, Runze Liu, Sa~Yang, Jingxuan Li, Xinwei Long, Jiaheng Ma, et~al.
\newblock Reviewrl: Towards automated scientific review with rl.
\newblock In {\em Proceedings of the 2025 Conference on Empirical Methods in Natural Language Processing}, pages 16942--16954, 2025.

\bibitem{weng2024cycleresearcher}
Yixuan Weng, Minjun Zhu, Guangsheng Bao, Hongbo Zhang, Jindong Wang, Yue Zhang, and Linyi Yang.
\newblock Cycleresearcher: Improving automated research via automated review.
\newblock {\em arXiv preprint arXiv:2411.00816}, 2024.

\bibitem{gao2024reviewer2}
Zhaolin Gao, Kiant{\'e} Brantley, and Thorsten Joachims.
\newblock Reviewer2: Optimizing review generation through prompt generation.
\newblock {\em arXiv preprint arXiv:2402.10886}, 2024.

\bibitem{yu2024automated}
Jianxiang Yu, Zichen Ding, Jiaqi Tan, Kangyang Luo, Zhenmin Weng, Chenghua Gong, Long Zeng, Renjing Cui, Chengcheng Han, Qiushi Sun, et~al.
\newblock Automated peer reviewing in paper sea: Standardization, evaluation, and analysis.
\newblock {\em arXiv preprint arXiv:2407.12857}, 2024.

\bibitem{zhuang2025large}
Zhenzhen Zhuang, Jiandong Chen, Hongfeng Xu, Yuwen Jiang, and Jialiang Lin.
\newblock Large language models for automated scholarly paper review: A survey.
\newblock {\em Information Fusion}, page 103332, 2025.

\bibitem{hossain2025llms}
Eftekhar Hossain, Sanjeev~Kumar Sinha, Naman Bansal, R~Alexander Knipper, Souvika Sarkar, John Salvador, Yash Mahajan, Sri Ram Pavan~Kumar Guttikonda, Mousumi Akter, Md~Mahadi Hassan, et~al.
\newblock Llms as meta-reviewers’ assistants: A case study.
\newblock In {\em Proceedings of the 2025 Conference of the Nations of the Americas Chapter of the Association for Computational Linguistics: Human Language Technologies (Volume 1: Long Papers)}, pages 7763--7803, 2025.

\bibitem{idahl2025openreviewer}
Maximilian Idahl and Zahra Ahmadi.
\newblock Openreviewer: A specialized large language model for generating critical scientific paper reviews.
\newblock In {\em Proceedings of the 2025 Conference of the Nations of the Americas Chapter of the Association for Computational Linguistics: Human Language Technologies (System Demonstrations)}, pages 550--562, 2025.

\bibitem{zhu2025deepreview}
Minjun Zhu, Yixuan Weng, Linyi Yang, and Yue Zhang.
\newblock Deepreview: Improving llm-based paper review with human-like deep thinking process.
\newblock {\em arXiv preprint arXiv:2503.08569}, 2025.

\bibitem{chang2025treereview}
Yuan Chang, Ziyue Li, Hengyuan Zhang, Yuanbo Kong, Yanru Wu, Hayden Kwok-Hay So, Zhijiang Guo, Liya Zhu, and Ngai Wong.
\newblock Treereview: A dynamic tree of questions framework for deep and efficient llm-based scientific peer review.
\newblock In {\em Proceedings of the 2025 Conference on Empirical Methods in Natural Language Processing}, pages 15662--15693, 2025.

\bibitem{taechoyotin2024mamorx}
Pawin Taechoyotin, Guanchao Wang, Tong Zeng, Bradley Sides, and Daniel Acuna.
\newblock Mamorx: Multi-agent multi-modal scientific review generation with external knowledge.
\newblock In {\em Neurips 2024 Workshop Foundation Models for Science: Progress, Opportunities, and Challenges}, 2024.

\bibitem{luagent}
Kai Lu, Shixiong Xu, Jinqiu Li, Kun Ding, and Gaofeng Meng.
\newblock Agent reviewers: Domain-specific multimodal agents with shared memory for paper review.
\newblock In {\em Forty-second International Conference on Machine Learning}.

\bibitem{anthropic_mcp2025}
{Anthropic}.
\newblock Introducing the model context protocol.
\newblock \url{https://www.anthropic.com/news/model-context-protocol}, 2024.
\newblock Accessed: 2025-11-09.

\bibitem{A2A_Protocol_2025}
Google.
\newblock Agent2agent (a2a) protocol — latest documentation.
\newblock \url{https://a2a-protocol.org/latest/}, 2025.
\newblock Accessed: 2025-11-09.

\bibitem{scp2025}
{Open Science Lab}.
\newblock Scientific intelligence context protocol (scp).
\newblock \url{https://github.com/open-sciencelab/scp}, 2025.
\newblock Accessed: 2025-11-09.

\bibitem{shi2024stochastic}
Zekun Shi, Zheyuan Hu, Min Lin, and Kenji Kawaguchi.
\newblock Stochastic taylor derivative estimator: Efficient amortization for arbitrary differential operators.
\newblock {\em Advances in Neural Information Processing Systems}, 37:122316--122353, 2024.

\bibitem{qu2025crispr}
Yuanhao Qu, Kaixuan Huang, Ming Yin, Kanghong Zhan, Dyllan Liu, Di~Yin, Henry~C Cousins, William~A Johnson, Xiaotong Wang, Mihir Shah, et~al.
\newblock Crispr-gpt for agentic automation of gene-editing experiments.
\newblock {\em Nature Biomedical Engineering}, pages 1--14, 2025.

\bibitem{swanson2025virtual}
Kyle Swanson, Wesley Wu, Nash~L Bulaong, John~E Pak, and James Zou.
\newblock The virtual lab of ai agents designs new sars-cov-2 nanobodies.
\newblock {\em Nature}, pages 1--3, 2025.

\bibitem{phan2025humanity}
Long Phan, Alice Gatti, Ziwen Han, Nathaniel Li, Josephina Hu, Hugh Zhang, Chen Bo~Calvin Zhang, Mohamed Shaaban, John Ling, Sean Shi, et~al.
\newblock Humanity's last exam.
\newblock {\em arXiv preprint arXiv:2501.14249}, 2025.

\bibitem{yao2023tree}
Shunyu Yao, Dian Yu, Jeffrey Zhao, Izhak Shafran, Tom Griffiths, Yuan Cao, and Karthik Narasimhan.
\newblock Tree of thoughts: Deliberate problem solving with large language models.
\newblock {\em Advances in neural information processing systems}, 36:11809--11822, 2023.

\bibitem{OpenAI2025GPT5}
{OpenAI}.
\newblock Gpt-5 is here, 2025.
\newblock Accessed: 2025-11-17.

\bibitem{du2025deepresearchbenchcomprehensivebenchmark}
Mingxuan Du, Benfeng Xu, Chiwei Zhu, Xiaorui Wang, and Zhendong Mao.
\newblock Deepresearch bench: A comprehensive benchmark for deep research agents, 2025.

\bibitem{Guo2024IdeaBenchBL}
Sikun Guo, Amir~Hassan Shariatmadari, Guangzhi Xiong, Albert Huang, Eric Xie, Stefan Bekiranov, and Aidong Zhang.
\newblock Ideabench: Benchmarking large language models for research idea generation.
\newblock {\em Proceedings of the 31st ACM SIGKDD Conference on Knowledge Discovery and Data Mining V.2}, 2024.

\bibitem{Ye2024JusticeOP}
Jiayi Ye, Yanbo Wang, Yue Huang, Dongping Chen, Qihui Zhang, Nuno Moniz, Tian Gao, Werner Geyer, Chao Huang, Pin-Yu Chen, Nitesh~V. Chawla, and Xiangliang Zhang.
\newblock Justice or prejudice? quantifying biases in llm-as-a-judge.
\newblock {\em ArXiv}, abs/2410.02736, 2024.

\bibitem{chen-etal-2024-humans}
Guiming~Hardy Chen, Shunian Chen, Ziche Liu, Feng Jiang, and Benyou Wang.
\newblock Humans or {LLM}s as the judge? a study on judgement bias.
\newblock In Yaser Al-Onaizan, Mohit Bansal, and Yun-Nung Chen, editors, {\em Proceedings of the 2024 Conference on Empirical Methods in Natural Language Processing}, pages 8301--8327, Miami, Florida, USA, November 2024. Association for Computational Linguistics.

\bibitem{lmarena2025}
{LMArena: Find the best AI for you}.
\newblock \url{https://lmarena.ai/}, 2025.
\newblock Accessed: 2025-11-07.

\end{thebibliography}

\end{document}